# A full-chain tube-based constitutive model

# for living linear polymers

J. D. Peterson and M. E. Cates

*University of Cambridge, Department of Applied Maths and Theoretical Physics, Centre for Mathematical Sciences, Wilberforce Road, Cambridge CB3 0WA*

Abstract:  We present a new strategy for introducing population balances into full-chain constitutive models of living polymers with linear chain architectures.  We provide equations to describe a range of stress relaxation processes covering both unentangled systems (Rouse-like motion) and well entangled systems (reptation, contour length fluctuations, chain retraction, and constraint release).  Special attention is given to the solutions that emerge when the 'breaking time' of the chain becomes fast compared to various stress relaxation processes.  In these 'fast breaking' limits, we reproduce previously known results (with some corrections) and also present new results for nonlinear stress relaxation dynamics.  Our analysis culminates with a fully developed constitutive model for the fast breaking regime in which stress relaxation is dominated by contour length fluctuations.  Linear and nonlinear rheology predictions of the model are presented and discussed.

## 1. Introduction

Living polymers are polymer-like materials whose constituent molecules can spontaneously break apart and recombine, leading to a dynamical (or 'living') equilibrium molecular weight distribution rather than a static one.  Over the years, living polymer systems of various chemistries (including wormlike micelles, liquid sulfur, and 'sticky' polymers) have been a subject of considerable experimental and theoretical interest [1] [2] [3] [4] [5] [6] [7] [8] [9].  Of all the chemistry platforms for living polymers, wormlike micelles (WLM) seem to have become the most popular system to study for rheologists.

To date, a great deal of progress has been made for understanding aspects of the linear rheology of well entangled living polymers; a Maxwellian response in the linear rheology can be attributed to the combined effects of reptation and reversible scission [1], and departures from said Maxwellian response can be used for making inferences to the scission energy [10] [3].  Progress on the nonlinear rheology of wormlike micelles is still an active area of research, with a great deal of attention given to 'shear banding' phenomena; one popular model suggests that shear banding is a result of flow-enhanced scission [4], but based on our recent work with the 'living Rolie Poly' model [11], this explanation now seems unlikely.  Other explanations for shear banding have included shear induced demixing [12] and isotropic/nematic phase transitions [5].  Shear induced phase transitions may be important for systems close to the boundary of a thermodynamic phase transition, but the simplest and most general mechanism for shear banding in any system of entangled linear polymers is that of the non-monotonic constitutive equation, as found in the Doi Edwards model [13] and engineering models like Johnson Segalman [14].

In the Doi Edwards model stress initially increases with increasing shear rate as tube segments become increasingly oriented in the axis of extension.  When the shear rate exceeds the rate of reptation, however, tube segments are rotated towards the flow direction, causing the shear stress to decrease with increasing shear rate. Under such conditions, homogeneous shear flow is linearly unstable to perturbations and shear banding can occur. The Doi Edwards instability is thought to be suppressed by 'convective constraint release' (CCR) in flexible unbreaking polymers [15], and some have argued that CCR is slightly weaker in living polymer systems due at least in part to a narrowing of the underlying stress relaxation spectra [16] [17]. If true, this could explain why shear banding is sometimes seen in WLM but not regularly seen in monodisperse unbreaking polymers. To more accurately assess such a hypothesis, however, a more detailed understanding is needed for the linear and nonlinear rheology of well entangled living polymers in the 'fast breaking' limit where shear banding in WLM is often found.

To that end, a final objective of the present paper is to introduce a new constitutive model for living polymers that accounts for all the usual modes of stress relaxation in well-entangled polymers (reptation, contour length fluctuations, chain retraction, constraint release), as well as the effects of reversible scission, in the fast breaking limit. Conceptually, this is not a new idea – the Reptation Reaction model [18] was built on similar principles, but omits constraint release terms because it was developed before the importance of constraint release was fully understood. Unfortunately, constraint release cannot be easily added on to the Reptation Reaction model – a different mathematical framework is required altogether.

Therefore, we begin our study by presenting a more complete mathematical framework for constitutive modelling of well-entangled living polymers. We apply this framework to consider linear stress relaxation processes in living polymers undergoing reptation, contour length fluctuations, chain retraction, and Rouse-like motion[1]. In the process, we produce new insights and a correction to a previously reported scaling result. Next, we show that this mathematical framework is easily generalized to nonlinear modes of stress relaxation, namely chain retraction and convective constraint release, and new results for 'fast breaking' modes of stress relaxation are put forward. Compiling these results together in view of a particular fast breaking regime (the 'fast breaking CLF' regime [1]) we obtain the STARM model (simplified tube approximation for rapid-breaking micelles), a new nonlinear constitutive model for well entangled living polymers that are prone to break apart and recombine on time-scales much faster than the time-scales for stress relaxation. It should be noted that our allusion to an assumed micellar chemistry in the name 'STARM' does not reflect a limitation to the range of chemistry platforms for which the model might be applied.

The STARM model, like the GLaMM model [15] and the MML model before it [16] is based on a full-chain tube-based model of stress relaxation dynamics. It is in the same family of constitutive equations as the popular single-mode Rolie Poly equation [19], and when constraint release relaxations are turned off, the STARM model reduces to the Rolie Poly equation. Overall, however, the STARM model is most closely related to the 'living polymer' version of the MML model [16]. The STARM model formalizes and sharpens the intuitions upon which that model was based and also offers an extension to flow conditions in which chains become stretched.

---

[1] Our analysis does not yet extend to stress relaxation processes relevant to highly stretched polymers (e.g. contours stretched to several times the equilibrium length). Under these conditions, one may need to think about flow-induced disentanglement (Mohageghi and Khomami, 2015), flow-induced scission [4], and flow-induced suppression of CCR [15]. It is our view that the modelling tools needed for this flow regime are still under development, and can be implemented in future versions of the STARM model.

Evaluating the linear and nonlinear rheology predictions of the STARM model, we find that constraint release (CR) processes seem to be equally effective for stress relaxation in living polymers and unbreaking polymers. Based on a comparison to experimental measurements, however, there is evidence that constraint release processes are strongly suppressed in wormlike micelles compared to what would be expected of a related system with flexible and unbreaking polymers. This points to one of two possible interpretations: first, it is possible that a reduced effectiveness of CCR emerges when the persistence length of the micelle (or polymer) begins to approach the length of an entanglement segment; second, it is possible that the standard 'continuum' description of Rouse like motion for constraint release relaxation is not suitable on time-scales comparable to or faster than the constraint release time.

We now summarize the outline of the present report. In section 2, we introduce a new mathematical framework for constitutive modelling of living polymers, which is then applied in sections 2.1 - 2.6 to study various modes of stress relaxation dynamics. In section 2.7, the results of the preceding sections are compiled together to form the STARM model. In section 3 we present a sampling of predictions from the STARM model in various linear and nonlinear rheometric flows. In section 4, we compare predictions of the STARM model to previously published experimental data. Finally, in section 5 we summarize the main results of our study and discuss directions of future research.

The discussion in sections 2.1 - 2.5 provides a core mathematical framework for constitutive modelling of living polymers, which is the primary purpose of the present report. However, readers who are mainly interested in the final assembly of a non-linear constitutive model (section 2.7) and its application (sections 3 and 4) can skip ahead. For these readers, we suggest that mathematical complexity of our report belies the simplicity of its core ideas, which are graphically shown in Figure 1 and Figure 2 (and later generalized in Figure 6 and Figure 7). We also recommend section 2.1.3 as an introduction to the supporting mathematical tools.

## 2. Derivation of the STARM model

In the current section, we present a new population balance framework for constitutive modelling of linear chain living polymers. This framework is sufficient to describe a wide range of systems, provided one has a suitable constitutive model for unbreaking chains undergoing the same set of relaxation processes. Here, we consider entangled systems undergoing reptation, reptation and contour length fluctuations (CLF), and chain retraction. We also consider unentangled systems undergoing Rouse-like motion. Our analysis re-affirms previously established results for reptation and contour length fluctuations, but yields new results for chain retraction and Rouse-like motions.

As a first application of this new framework, we have focused on systems that are 'fast breaking', where chains break apart and combine with other chains on timescales much faster than the typical chain's longest relaxation time in the absence of any such reactions. This is a limit that is believed to describe some formulations of well entangled wormlike micelles [5] [20]. Our analysis of stress relaxation in the fast breaking CLF limit culminates in a new nonlinear constitutive equation, the STARM model.

### 2.1. Reptation Revisited

For well-entangled linear living polymers, it is well known that when the typical breaking time for a chain, $\tau_B$, is much faster than the typical reptation time in the absence of breaking, $\tau_{rep}$, the material relaxes its stress with a single relaxation time $\tau \sim \tau_{rep}\zeta^{1/2}$, where $\zeta = \tau_B/\tau_{rep}$ [1]. Physically, we interpret these results as follows:

With regard to the terminal relaxation time, $\tau \sim \tau_{rep}\zeta^{1/2}$: when a break first occurs along a chain, only the tube segments very near to the breakpoint will relax their stress before the chain ends recombine with other chain ends. For a chain of average length, each break relaxes a fraction $\sim \zeta^{1/2}$ of the chain and it takes a time $\tau \sim \tau_B/\zeta^{1/2} = \tau_{rep}\zeta^{1/2}$ for the whole chain to relax its stress.

With regard to the single relaxation time: whereas most of the stress relaxation takes place at chain ends, most of the stress is still held by segments within the chain interior. Moreover, scission and recombination ensure that the mean stress held by an interior segment is the same for all molecular weights and all contour positions (except those very close to the chain end). While shorter chains may relax their stress faster than longer chains, the rate of stress relaxation for any chain is always proportional to the stress held by its interior segments. Thus, the overall rate of stress relaxation is (to leading order) proportional to the overall stress, and the system exhibits a single relaxation time.

An equivalent (and possibly more familiar) interpretation of the single relaxation time can also be given: when reactions quickly update the length of chain into which a particular entanglement segment has been incorporated, chain segments are able to explore the entire molecular weight distribution before relaxing any stress. Likewise there is no dependence on position within the chain, whereas for unbreakable polymers the relaxation rate is faster near the (permanent) chain ends. Thus, the relaxation spectra for all tube segments must be the same even though the molecular weight distribution is polydisperse at any given moment in time [1].

The results for reptation in living polymers are already well-established in the literature [1] [17], and so they serve as a useful starting point for introducing and validating our more complete mathematical framework for describing stress relaxation in living polymer systems. Along the way, we will also present some new results to describe how different sectors of the molecular weight distribution contribute to stress relaxation.

### 2.1.1. Governing Equations

In well entangled unbreaking polymers, the process of stress relaxation by reptation (in the absence of any thermal constraint release) can be modelled via a 1D diffusion equation. Assuming the stress held by a tube segment only relaxes when the occupying chain diffuses past said tube segment (and no portion of the chain remains within), we can equate stress relaxation with the 'death' of tube segments by curvilinear diffusion. The probability $P(t,s)$ that a tube segment initially at contour position $s \in [0, Z]$ along the chain is still 'surviving' at time $t$ evolves by[2]:

$$\frac{\partial P}{\partial t} = D_C \frac{\partial^2 P}{\partial s^2} \qquad (1)$$

If all tube segments are initially unrelaxed and chain ends are always stress free, we assign an initial condition $P(t = 0, s) = 1$ and boundary conditions $P(t, 0) = P(t, Z) = 0$. Note that the contour position $s$ is indexed in units of entanglement segments, so that $Z$ is the entanglement number of the chain. Here, as elsewhere in the report, the entanglement number $Z$ refers to the equilibrium entanglement number and is assumed to be linear in the molecular weight. The curvilinear diffusion coefficient $D_C$ is related to the entanglement number $Z$ and the Rouse time of an entanglement segment $\tau_e$ by $D_C = 1/3\pi^2 Z \tau_e$ [15].

---

[2] In a full-chain tube-based model including constraint release dynamics, stress relaxation dynamics are written in terms of a tangent correlation tensor instead of a tube survival probability (c.f. section 2.4 - 2.7). When constraint release is absent, equations (1) - (2) will naturally emerge from the linear rheology but the equations more precisely describe a relaxation of alignment rather than the survival of tube segments.

The fraction of unrelaxed stress is proportional to the mean tube survival probability, $\bar{P}$, such that the stress relaxation modulus $G(t)$ is given by:

$$\frac{G(t)}{G_e} = \bar{P}(t) = \frac{1}{Z} \int_0^Z ds\, P(t,s) \qquad (2)$$

where $G_e$ is the plateau modulus of the entangled melt. If the system of interest contains a polydisperse mixture defined by a number density distribution $n(Z)$[3], then for each chain length $Z$, the tube survival probability $P(t,s,Z)$ for a chain of with $Z$ entanglements is given by:

$$\frac{\partial}{\partial t} P(t,s,Z) = D_C(Z) \frac{\partial^2}{\partial s^2} P(s,t,Z) \qquad (3)$$

and the stress relaxation modulus includes an integral on $Z$:

$$\frac{G(t)}{G_e} = \bar{P}(t) = \int_0^\infty dZ'\, n(Z') \int_0^{Z'} ds\, P(t,s,Z') \Big/ \int_0^\infty n(Z')Z'dZ' \qquad (4)$$

In general, a portion of the molecular weight distribution may lie below the entanglement threshold, $Z < 1$, but we will not bother discussing a separate treatment of stress relaxation for such chains. Instead, we assume that when most chains are well entangled, the details of stress relaxation for the small fraction of unentangled chains is comparatively unimportant.

If the polydisperse mixture of chains are also undergoing reversible scission reactions, then one must perform a detailed accounting (population balances) to describe changes in both the molecular weight distribution and the distribution of surviving tube segments. Using elementary reaction kinetics, the number density $n(Z)$ of chains with entanglement number $Z$ evolves according to [1]:

$$\frac{\partial}{\partial t} n(Z) = -c_1 \ell_e n(Z)Z - 4c_2 \int_0^\infty dZ'n(Z)n(Z') + 2c_2 \int_0^Z dZ'n(Z')n(Z-Z')$$
$$+ 2c_1 \ell_e \int_Z^\infty dZ'n(Z') \qquad (5)$$

where $c_1$ is the rate constant for scission (units of $1/\ell_P$ per unit time, where $\ell_P$ is the persistence length of the micelle), $\ell_e$ is the length of an entanglement strand, and $c_2$ is the rate constant for recombination (units of volume per free chain end per unit time). In addition to reversible scission, there are other reversible polymer reactions to consider. For example, the free end of one chain may attach itself to the backbone of a nearby chain, forming a temporary branch point. When this temporary branch point breaks apart, the two product chains can be different from the original reactants. This reaction pathway, with rate constant $c_3$, is known as 'end attack' (also called 'end interchange') and it causes the number density distribution to evolve as:

$$\frac{\partial}{\partial t} n(Z) = \cdots + 2c_3\ell_e \left[ -n(Z)Z \int_0^\infty dZ' n(Z') - n(Z) \int_0^\infty dZ' n(Z') Z' + \int_0^Z dZ' \int_{Z-Z'}^\infty dZ'' \left[ n(Z')n(Z'') \right] \right.$$
$$\left. + \left[ \int_0^\infty dZ' n(Z') \right] \left[ \int_Z^\infty dZ' n(Z') \right] \right] \qquad (6)$$

---

[3] A continuous number density distrubtion $n(Z)$ means that the number density of chains in the interval $Z' \in [Z, Z+\delta Z]$ is given by $n(Z)\delta Z$ in the limit of $\delta Z \to 0$. The molecular weight of a chain is a discrete variable in reality, but we can treat it as continuous when the monomer unit (a single surfactant molecule) is very much smaller than the size of an entanglement segment.

For the remainder of the present report, we will assume that the molecular weight distribution $n(Z)$ is the equilibrium molecular weight distribution given a total number density of persistence lengths, $\rho$:

$$n(Z) = n_0 \exp\left(-\frac{Z}{\bar{Z}}\right) \qquad \bar{Z}\ell_e = \left[2\rho\ell_P \frac{c_2}{c_1}\right]^{1/2} \qquad n_0 = \frac{1}{2}\frac{c_1\ell_e}{c_2} \qquad (7)$$

where $\bar{Z}$ is an average entanglement number for the melt. Previous work with the 'living Rolie Poly' model has shown that couplings between stress and reaction kinetics can be safely ignored whenever the stretching energy in an entanglement segment is small compared to the scission energy [11]. If the typical scission energy is on the order of $10k_BT$, this restriction is most concerning for steady uni-axial extension when the strain rate exceeds the terminal stretch relaxation time. For lower strain rates and for simple shear flows, flow-induced changes to the molecular weight distribution are less concerning.

There are other reaction processes that one may consider (e.g. bond interchange and end evaportation) but for the time being we limit our focus to reversible scission and end attack, as they are regarded as the two pathways most likely to contribute to stress relaxation in most living polymer systems [21].

Whenever reactions transport an entanglement segment from one length population to another, the stress associated with that entanglement segment is transported as well. Precisely stated, if two chains with entanglement numbers $Z_1$ and $Z_2$ combine to form a chain of length $Z_3 = Z_1 + Z_2$, then given tube survival profiles $P_1(s)$ and $P_2(s)$ for the two shorter chains, the tube survival profile of the longer chain will be:

$$P_3(s) = \begin{cases} P_1(s) & \text{if } s < Z_1 \\ P_2(s-Z_1) & \text{if } s \geq Z_1 \end{cases} \qquad (8)$$

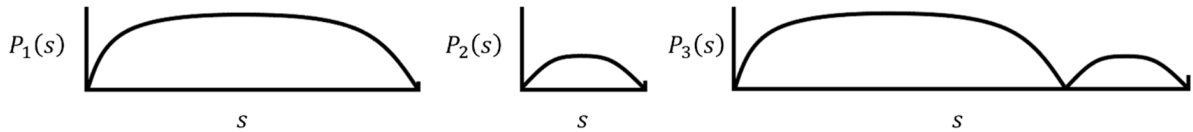

*Figure 1: Cartoon of the attachment rule, equation (8)*

Equation (8) assumes that whenever two chains become attached, the time-scale over which the transition occurs is extremely slow compared to the timescales over which chains relax their stress or change their configuration. Likewise, if a chain of entanglement number $Z_3$ breaks at $s = Z_1$, then it will form two chains with entanglement numbers $Z_1$ and $Z_2 = Z_3 - Z_1$. If the tube survival profile of the original long chain is $P_3(s)$, then the tube survival profile of the fragment chains will be:

$$P_1(s) = P_3(s) \qquad P_2(s) = P_3(s - Z_1) \qquad (9)$$

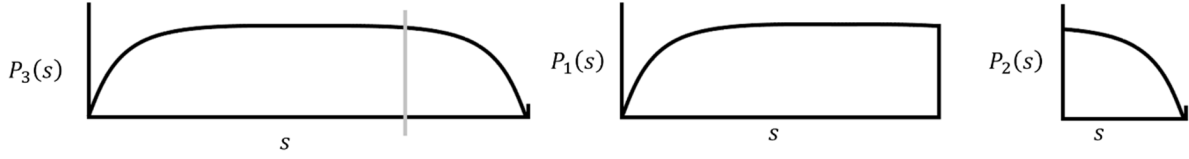

*Figure 2: Cartoon of the scission rule, equation (9).*

Note that *at the moment a reaction occurs*, the mean tube survival probability is conserved by both scission and recombination. In this view, scission and recombination cannot inherently relax any stress; they can only contribute to a speed-up of underlying stress relaxation processes such as reptation.

When scission and recombination are repeated many times over, equations (8) and (9) describe a 'shuffling' of surviving tube segments through the molecular weight distribution. Following [1], we will assume that a pair of recently formed chain ends are more likely to find new partners than recombine, and so the 'shuffling' effect can be accounted for directly in the tube survival probability equation. For a reaction scheme dominated by reversible scission, we write:

$$\frac{\partial}{\partial t}\big(n(Z)P(s,Z)\big) = D_C(Z)n(Z)\frac{\partial^2}{\partial s^2}P(s,Z) - c_1\ell_e P(s,Z)n(Z)Z$$

$$+ c_1\ell_e \int_Z^\infty dZ' n(Z')\big(P(s,Z') + P(Z'-s,Z')\big)$$

$$- c_2 P(s,Z)\int_0^\infty dZ' n(Z)n(Z')$$

$$+ c_2 \int_0^Z dZ' n(Z')n(Z-Z') \Bigg[ \begin{cases} P(s,Z') & \text{if } s < Z' \\ P_2(s-Z',Z-Z') & \text{if } s \geq Z' \end{cases} \Bigg\}$$

$$+ \begin{cases} P(s,Z-Z') & \text{if } s < Z' \\ P(s-Z+Z',Z') & \text{if } s \geq Z' \end{cases} \Bigg] \qquad (10)$$

The reaction terms are written in parallel to the reaction terms in equation (5). The first two terms represents the loss of unrelaxed tube segments in chains of length $Z$ due to breaking/addition events, after which the original tube segments are still surviving but incorporated into chains of smaller/larger lengths. The third and fourth terms represent the addition of unrelaxed tube segments in chains of length $Z$ due to the formation of chains with length $Z$ by addition of shorter chains and scission of longer chains, respectively.

A parallel set of population balance equations can be written when the end attack reaction pathway is dominant:

$$\frac{\partial}{\partial t}\big(n(Z)P(s,Z)\big) = D_C(Z)n(Z)\frac{\partial^2}{\partial s^2}P(s,Z) - 2c_3\ell_e P(s,Z)n(Z)Z\int_0^\infty dZ' n(Z')$$

$$-2c_3\ell_e P(s,Z)n(Z)\int_0^\infty dZ'n(Z')Z'$$

$$+c_3\ell_e\left[\int_0^\infty dZ'n(Z')\right]\left[\int_Z^\infty dZ'n(Z')\big(P(s,Z')+P(Z'-s,Z')\big)\right]$$

$$+c_3\ell_e\int_0^Z dZ'\int_{Z-Z'}^\infty dZ''\,n(Z')n(Z'')\left[\begin{cases}P(s,Z') & \text{if } s<Z'\\ P_2(s-Z+Z',Z'') & \text{if } s\ge Z'\end{cases}\right.$$

$$\left.+\begin{cases}P(s,Z'') & \text{if } s<Z-Z'\\ P(s-Z+Z',Z') & \text{if } s\ge Z-Z'\end{cases}\right] \qquad (11)$$

By visual inspection, it is clear that the simplicity of equations (8) and (9) belies the complexity that reversible scission adds to stress relaxation dynamics in living polymer systems. To that end, we also present a 'toy' depiction of the population balance terms, similar to the 'mean field' approximation of Lequeux [2] and to the Poisson Renewal model of Granek and Cates [10]. Ignoring the detailed accounting of equation (10), one can approximate that on a time-scale $\tau_{sh}$, every tube segment loses all memory of past associations and can be found at any contour position on a chain of any length with equal likelihood. In other words, we assume that reactions cause tube segments to become uniformly shuffled (or re-distributed) through the molecular weight distribution on a timescale $\tau_{sh}$:

$$\frac{\partial}{\partial t}\big(n(Z)P(t,s,Z)\big)=D_C(Z)n(Z)\frac{\partial^2}{\partial s^2}P(s,t,Z)-\frac{1}{\tau_{sh}}n(Z)(P-\bar{P}) \qquad (12)$$

This 'mean field' approximation (12) of the full reversible scission equation (10) captures two important features of the full model. First, we see that after integrating over the whole molecular weight distribution, the 'shuffling' term has no direct contribution to changing the mean tube survival probability. Second, scission and recombination tend to homogeneize the tube survival probability; whenever a long chain (more surviving tube segments) breaks, the surviving tube segments in the fragment chains increase the average tube survival probability in their respective length populations. Likewise, whenever two short chains (fewer surviving tube segments) combine, the resulting long chain decreases the mean tube survival probability in its length population. Finally, it goes without saying that the 'toy' shuffling model becomes equivalent to the reversible scission or end attack model in the limit where reaction terms are switched off ($\tau_{sh}\to\infty$).

### 2.1.2. Governing Equations (Dimensionless)

Before proceeding with any calculations, we find it helpful to non-dimensionalize. In the absence of reactions, tube segments relax on a timescale comparable to the reptation time for a chain of average length, $\tau_{rep}=\bar{Z}^2/D_C(\bar{Z})$, and changes in $P$ develop over contour lengths comparable to the mean entanglement number $\bar{Z}$. Thus, we define a dimensionless time $\bar{t}=t/\tau_{rep}$ and a rescaled contour index $\bar{s}=s/\bar{Z}$. However, for visual clarity we omit overbars in the equations that follow – for the remainder of this section, all variables are assumed to be dimensionless unless otherwise stated. We also continue to assume the equilibrium molecular weight distribution (7).

The dimensionless form of the toy shuffling model (12) is given by:

$$\zeta_{sh}\frac{\partial}{\partial t}P = \zeta_{sh}\frac{1}{z}\frac{\partial^2 P}{\partial s^2} - (P - \bar{P}) \qquad (13)$$

where $z = Z/\bar{Z}$ and $\zeta_{sh} = \tau_{sh}/\tau_{rep}$. The dimensionless equations for reversible scission are:

$$\zeta_{RS}\frac{\partial}{\partial t}P = \zeta_{RS}\frac{1}{z}\frac{\partial^2 P}{\partial s^2} - Pz$$

$$+ e^z\int_z^\infty dz' e^{-z'}\big(P(s,z') + P(z'-s,z')\big)$$

$$-2P(s,z)$$

$$+\frac{1}{2}\int_0^z dz'\left[\begin{cases} P(s,z') & \text{if } s < z' \\ P(s-z',z-z') & \text{if } s \geq z' \end{cases}\right\} + \left\{\begin{cases} P(s,z-z') & \text{if } s < z' \\ P(s-z+z',z') & \text{if } s \geq z' \end{cases}\right] \qquad (14)$$

where $\zeta_{RS} = \tau_B^{RS}/\tau_{rep}$ and $\tau_B^{RS} = [c_1\bar{Z}\ell_e]^{-1}$ is the breaking time for a chain of average length undergoing reversible scission. Finally, the dimensionless equations when end attack is dominant are given by:

$$\zeta_{EA}\frac{\partial}{\partial t}P = \zeta_{EA}\frac{1}{z}\frac{\partial^2 P}{\partial s^2} - 2P(s,z)z - 2P(s,Z)$$

$$+ e^z\left[\int_z^\infty dz' e^{-z'}\big(P(s,z') + P(z'-s,z')\big)\right]$$

$$+ e^z\int_0^z dz'\int_{z-z'}^\infty dz''\, e^{-z'-z''}\left[\begin{cases} P(s,z') & \text{if } s < z' \\ P(s-z+z',z'') & \text{if } s \geq z' \end{cases}\right\}$$

$$+ \left\{\begin{cases} P(s,z'') & \text{if } s < Z-Z' \\ P(s-z+z',z') & \text{if } s \geq z-z' \end{cases}\right] \qquad (15)$$

where $\zeta_{EA} = \tau_B^{EA}/\tau_{rep}$ and $\tau_B^{EA} = [c_3 n_0\bar{Z}\ell_e]^{-1}$ is the breaking time for a chain of average length undergoing end attack.

At this point, we can pursue a solution to the governing equations in the 'fast breaking' limit, where the typical chain is expected to break many times before it is able to completely relax its stress by reptation. Before proceeding to solutions of the full models, it is instructive to begin with the 'toy' model.

### 2.1.3. The fast breaking limit: 'toy' calculations

In the 'fast breaking' limit, $\zeta_{sh} \ll 1$, the 'shuffling' term in equation (13) dominates over the other terms so that on time-scales comparable to reptation, the tube survival probability is homogenized across the molecular weight distribution and across all possible chain contour positions:

$$P(t,s,z) = \bar{P}(t) \qquad (16)$$

However, this leading order solution is incompatible with the boundary condition of stress free chain ends, $P(t,s = 0,z) = 0$. While it may be true that shuffling dominates over diffusion for all interior segments of a chain, sufficiently close to the chain ends it must be true that diffusion is faster than shuffling. To see how diffusion and shuffling balance very near to a chain end in the fast breaking limit, we rescale our chain contour coordinate to $\tilde{s} = s/\zeta_{sh}^{1/2}$ so that equation (13) with $\zeta_{sh} \ll 1$ becomes:

$$\frac{1}{z}\frac{\partial^2 P}{\partial \tilde{s}^2} - P + \bar{P} = 0 \qquad (17)$$

with boundary conditions that ensure stress free ends and matching to the 'outer' solution:

$$P(\tilde{s} = 0) = 0 \qquad P(\tilde{s} \to \infty) = \bar{P}$$

Equation (17) can be integrated to obtain:

$$P = \bar{P}\big[1 - \exp(-\tilde{s}\sqrt{z})\big] \qquad (18)$$

Now integrating equation (13) over the entire molecular weight distribution we find:

$$\frac{\partial \bar{P}}{\partial t} = 2 \int_0^\infty dz \left[\frac{e^{-z}}{z}\frac{\partial P}{\partial s}\right] \qquad (19)$$

and using equation (18), this becomes:

$$\frac{\partial \bar{P}}{\partial t} = -\bar{P}\zeta_{sh}^{-1/2}\left[2\int_0^\infty dz \frac{e^{-z}}{\sqrt{z}}\right] = -\left[\frac{4\pi}{\zeta_{sh}}\right]^{1/2}\bar{P} \qquad (20)$$

Thus, we see the usual single-exponential decay in $\bar{P}$ with a relaxation time (in units of $\tau_{rep}$) of $\tau = 0.28\zeta_{sh}^{1/2}$. The same scaling law for the overall relaxation time has been derived many times over, relying on the same underlying physical intuitions [1] [10]. However, we will soon show that the set of tools employed here (boundary layer methods for singular asymptotic problems) are powerful enough to make progress on nonlinear aspects of living polymer rheology where intuition alone might fail.

The 'toy' shuffling model is also useful for making calculations when $\zeta_{sh}$ is not very small, and we refer the interested reader to the appendix for additional details. Living polymers outside the fast breaking limit will be the particular focus of a follow-up paper to the present work.

### 2.1.4. The fast breaking limit: reversible scission and end attack

Here, we repeat the analysis of the preceding section for the full population balance equations describing the tube survival probability in living polymer systems undergoing reversible scission and end attack reactions.

In the fast breaking limit, $\zeta_{RS}, \zeta_{EA} \ll 1$, the leading order reversible scission and end attack equations prescribe $P(t, s, z) = \bar{P}(t)$, just as was found for the 'toy' model: when the reactions are fast compared to stress relaxation, we expect that the tube survival probability will be homogenized throughout the molecular weight distribution. Rescaling to consider stress relaxation very near the chain ends, we find for reversible scission ($\tilde{s} = s/\zeta_{RS}^{1/2}$):

$$0 = \frac{1}{z}\frac{\partial^2 P}{\partial \tilde{s}^2} - (P - \bar{P}) + \left[\int_z^\infty dz' e^{z-z'}P(\tilde{s}, z') - P\right] + \left[\int_0^z dz' P(s, z') - zP\right] \qquad (21)$$

And for end attack ($\bar{s} = s/\zeta_{EA}^{1/2}$):

$$0 = \frac{1}{z}\frac{\partial^2 P}{\partial \bar{s}^2} - (P - \bar{P}) - \left[\int_z^\infty dz' e^{z-z'} P(\bar{s}, z') - P\right] - \left[\int_0^z dz' P(s, z') - zP\right]$$
$$- \left[\int_0^z dz' n(z') \int_{z-z'}^\infty dz'' n(z'') P(s, z'') - zP\right] \quad (22)$$

The fast breaking equations for end attack and reversible scission differ only with respect to the last term in the end attack equations, which represents loss of an existing chain via end attack from another chain vs the gain via newly-joined segments from an end attack. However, for the shortest chains (which have a disproportionate contribution to the total stress relaxation) this term appears to be relatively unimportant – by virtue of their length, short chains are difficult targets and unlikely products (newly-joined) for end attack.

We can solve these fast breaking equations numerically to obtain $P(\bar{s}, z)/\bar{P}$. To compare the predictions of equations (18), (21), and (22) we plot contours of $P/\bar{P} = 1/2$ in Figure 3, marking the contour position $\bar{s}$ at which a tube segment on a chain of size $z$ is half as likely to be unrelaxed compared to tube segments deep in the chain interior. We see that for chains of shorter-than-average length, there is no discernible difference between reversible scission and end attack. The 'toy' model, by comparison, seems to slightly overestimate the stress relaxation occurring in short chains and also underestimate the number of unrelaxed end segments held by the longest chains. In units of $\tau_{rep}$, the relaxation times for fast breaking reptation with reversible scission and end attack are $0.26\zeta_{RS}^{1/2}$ and $0.27\zeta_{EA}^{1/2}$, respectively. The 'toy' model prediction of $0.28\zeta_{sh}^{1/2}$ is in good agreement with both results if the shuffling time is equated to the breaking time.

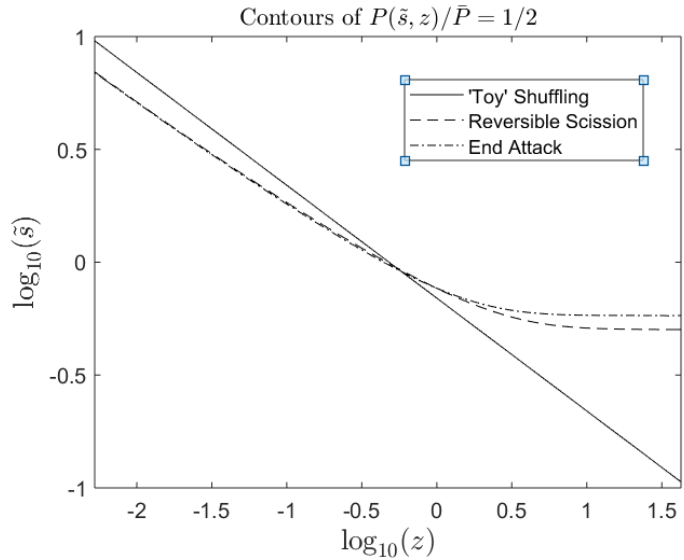

Figure 3: *Solution to the reversible scission and end attack equations in the fast breaking limit. Note that tube segments are most likely to be relaxed very near to the chain end, $\bar{s} = 0$, and less likely to be relaxed in the chain interior, $\bar{s} \to \infty$. Additionally, shorter chains have more relaxed tube segments at their ends than longer chains.*

In Figure 3 , we see that under reversible scission or end attack, long chains tend to have more relaxed ends than what would be predicted by the 'toy' approximation of the reaction scheme. In this case, the more relaxed end segments do not mean that the 'toy' model underestimates stress relaxation occurring in the longest chains. Instead, relaxed tube segments appear at the end of long chains because long chains of a given length $z > 1$ are more likely to arise by combination of shorter chains than by scission of even longer chains – the end segments of a long chain look similar to the end segments of much shorter chains simply because those end segments were in fact (a short time earlier) the end segments of a much shorter chain. By contrast, the 'shuffling' approximation continually refreshes all chains at all contour positions (including long chain ends) without regard for details of the underlying reactions.

Compared to the 'full' models for reversible scission and end attack, the 'toy' model underestimates the number of relaxed tube segments at the ends of a long chain and therefore overestimates the amount of tube segments that relax due to reptation of long chains. Fortunately, the population of chains for which the 'toy' model visibly fails (e.g. $z > 2$) only accounts for about 5% of the total stress relaxation (c.f. equation (20)). Therefore, significant differences in the predictions for long chains do not translate to significant differences in the final picture for stress relaxation.

Overall, we find that all three reaction schemes (reversible scission, end attack, and 'toy' shuffling) lead to very similar descriptions of stress relaxation in fast breaking living polymers. Therefore, for the remainder of this report, we will assume a reversible scission reaction scheme and employ a single effective value of the breaking time, $\tau_B$, which we define as:

$$\frac{1}{\tau_B} = \frac{1}{\tau_B^{RS}} + \frac{1}{\tau_B^{EA}} \qquad (23)$$

Or equivalently, with $\zeta = \tau_B/\tau_{rep}$:

$$\zeta^{-1} = \zeta_{EA}^{-1} + \zeta_{RS}^{-1} \qquad (24)$$

Likewise, a suitable estimate for the effective 'shuffling' time in the fast breaking limit is simply $\tau_{sh} = \tau_B$, or $\zeta_{sh} = \zeta$.

This concludes our discussion of fast breaking living polymers relaxing their stress by reptation alone. Besides reptation, there are many other ways in which linear chain living polymers might relax their stress, including contour length fluctuations, chain retraction, and constraint release. In the sections that follow, we will employ the same constitutive modelling tools to build up a more complete description of stress relaxation dynamics in fast breaking living polymers.

## 2.2. Reptation and Contour Length Fluctuations

In a system of unbreaking polymers, the tube survival equation (3) can be modified to include stress relaxation by contour length fluctuations. When the contour length of a chain is allowed to fluctuate (e.g. via thermal fluctuations) the end-most segments of the chain can escape their confining tube segments on timescales much faster than what would otherwise be predicted by reptation alone [22]. Once again, this is a 'linear' mode of stress relaxation and we are not yet interested in the system's response to large deformations.

Following the ideas of Likhtman and McLeish [23] [15], we approximate the effects of CLF via a second, $s$-dependent contribution to the diffusion coefficient in equation (3). In particular, the added

CLF term allows stress relaxation to occur more quickly for tube segments very near to a chain end, $s = 0, z$. In our dimensionless units, we write:

$$\frac{\partial P}{\partial t} = \frac{\partial}{\partial s}\left[\left(\frac{1}{z} + \frac{1}{\bar{Z}}\frac{1}{s^2(z-s)^2}\right)\frac{\partial P}{\partial s}\right] \qquad (25)$$

Recalling that $s$ and $z$ are both scaled by $\bar{Z}$, we find that CLF is dominant over reptation whenever:

$$\frac{1}{\bar{Z}}\frac{1}{s^2} > \frac{1}{z} \qquad (26)$$

$$s\bar{Z} < \sqrt{z\bar{Z}} \qquad (27)$$

Therefore, this model captures the fact that CLF dominates over reptation whenever a tube segment is within a distance of order $\sqrt{Z}$ tube segments from the chain end. Note also that whereas the diffusion coefficient for reptation is $z$-dependent, the additional term for CLF is not.

Adding in reactions (reversible scission) and rescaling to consider stress relaxation very near the chain ends, we obtain ($\tilde{s} = s\zeta^{-1/2}$):

$$0 = \frac{\partial}{\partial \tilde{s}}\left[\left(\frac{1}{z} + 3\frac{\zeta_R^{-1}}{\tilde{s}^2}\right)\frac{\partial P}{\partial \tilde{s}}\right] - (P - \bar{P}) + \left[\int_z^\infty dz' e^{z-z'}P(\tilde{s}, z') - P\right] + \left[\int_0^z dz' P(s, z') - zP\right] \qquad (28)$$

where $\zeta_R = 3\bar{Z}\zeta = \tau_B/\tau_R$ is the ratio of the breaking time to the longest stretch relaxation time or Rouse time, $\tau_R = \bar{Z}^2\tau_e$, for a chain with $\bar{Z}$ entanglement segments. Note that tube-based depiction of stress relaxation breaks down if the breaking time falls below the entanglement time, $\tau_B < \tau_e$, or equivalently, $\zeta_R\bar{Z}^2 < 1$.

When contour length fluctuations dominate over reptation, which we call the fast breaking CLF limit $\zeta_R \ll 1$, equation (28) has an analytic solution in terms of the modified Bessel functions $I_{3/4}(x)$ and $I_{-3/4}(x)$ ($\tilde{s} = s\zeta^{-1/4}\bar{Z}^{1/4}$):

$$\frac{P(\tilde{s}, z)}{\bar{P}} = \frac{P(\tilde{s})}{\bar{P}} = 1 - \frac{\Gamma(1/4)}{2\sqrt{2}}\left(I_{3/4}\left(\frac{1}{2}\tilde{s}^2\right) - I_{-3/4}\left(\frac{1}{2}\tilde{s}^2\right)\right) \qquad (29)$$

From this we find that for $\zeta_R \ll 1$ the stress relaxes on a single timescale (in units of $\tau_{rep}$) given by:

$$\tau = \zeta^{1/2}\zeta_R^{1/4}\left[\frac{2^{3/2}3^{1/4}}{\pi}\Gamma\left(\frac{1}{4}\right)\Gamma\left(\frac{5}{4}\right)\right]^{-1} \approx 0.26\zeta^{1/2}\zeta_R^{1/4} \qquad (30)$$

This result is in agreement with scaling arguments first proposed by Cates [1].

For values of $\zeta_R$ that are neither very small nor very large, we can solve equation (28) numerically. In Figure 4, we present solutions to equation (28) for $P/\bar{P} = 1/2$ in a system with $\zeta_R = 10^{-2}, 10^0, 10^2$ and $\zeta \ll 1$. The contour position $\tilde{s}$ has been scaled per equation (28), $\tilde{s} = s/\zeta^{1/2}$. These calculations represent systems that are well within the 'fast breaking' limit $\zeta \ll 1$ but transitioning from the fast breaking reptation limit, $\zeta_R \gg 1$ to the fast breaking CLF limit, $\zeta_R \ll 1$. For chains shorter than $z \sim \zeta_R^{1/2}$, reptation dominates over CLF and the contour of $P/\bar{P} = 1/2$ collapses to the result for $\zeta_R \to \infty$. For $\zeta_R < 1$, we also see that the number of unrelaxed end segments at the end of a typical chain increases with decreasing $\zeta_R$, with the threshold in $\bar{s}$ for $P/\bar{P} = 1/2$ scaling as $\bar{s} \sim \zeta_R^{1/4}$.

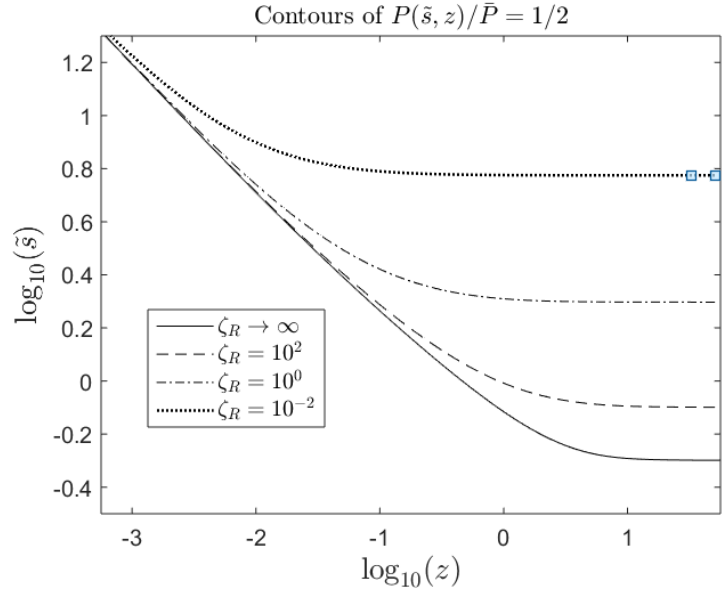

*Figure 4: Solutions to $P(\bar{s}, z)/\bar{P}$ for a living polymer system undergoing reversible scission, reptation, and CLF in the fast breaking limit, as described by equation (28). Note that for increasing $\zeta_R$, the solution transitions from CLF dominated (end segments uniformly relaxed at all z) to reptation dominated with shorter chains having more end segments relaxed than longer chains.*

Finally, decreasing $\zeta_R$ reduces the stress relaxation time, $\tau$. For $\zeta_R \to \infty$, the stress relaxation time (in units of $\tau_{rep}$) approaches $\tau = 0.26\zeta^{1/2}$ and for $\zeta_R \to 0$, the stress relaxation time approaches $\tau = 0.26\zeta^{1/2}\zeta_R^{1/4}$. In Figure 5, we show that there is a smooth transition between these two scaling regimes centered about $\zeta_R \sim 1$:

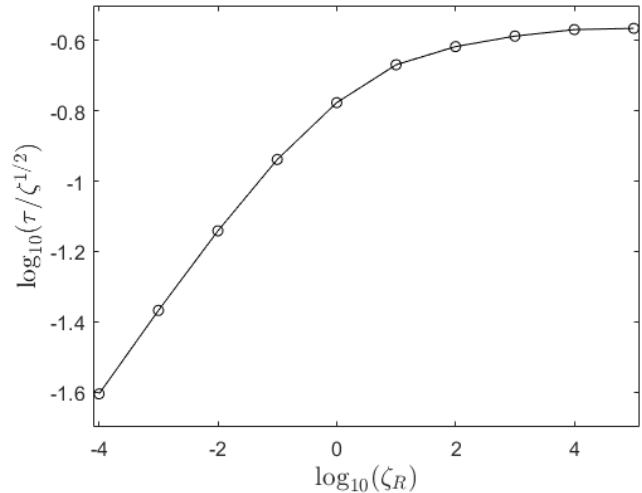

*Figure 5: Evaluating the stress relaxation time from equation (28) $\tau$ over a range of conditions within the fast breaking limit, $\zeta \ll 1$, varying from CLF dominated $\zeta_R \ll 1$ to reptation dominated $\zeta_R \gg 1$.*

The transition from slow breaking to fast breaking comes with a narrowing of the orientational relaxation spectra to just a single timescale, but there is no signature in the orientational relaxation spectra to mark the transition from fast breaking reptation to fast breaking CLF. However, the transition to $\zeta_R \ll 1$ does bring about significant changes in the relaxation spectra for chain stretching.

## 2.3. Chain retraction (Stretch relaxation only)

Here, we momentarily shift our focus away from a chain escaping its confining tube. Instead, we focus on the subject of a stretched chain retracting within its tube. If the stretched chain is modelled as a bead/spring assembly with linear springs and linear hydrodynamic drag, then in the limit where variations in stretch span many individual bead/spring segments, the local degree of stretching along the chain $\lambda(s)$ evolves according to [24]:

$$\frac{\partial \lambda}{\partial t} = -\frac{1}{\tau_R}\frac{\partial^2}{\partial s^2}(\lambda - 1) \qquad (31)$$

where $\tau_R = Z^2 \tau_e$ is the longest Rouse relaxation time of the chain and $s$ is normalized by the chain's entanglement number, $Z$. Note that we have defined $\lambda = 1$ as the unstretched state, so the constraint of stretch-free chain ends is given by:

$$\lambda(t, s = 0,1) = 1 \qquad (32)$$

Finally, we assume a uniform stretch as the initial condition:

$$\lambda(t = 0, s) = \lambda_0 \qquad (33)$$

When chain retraction is viewed only as a process for stretch relaxation, we see that it is a linear relaxation process. For nonlinear deformations, however, chain retraction effects the alignment of entanglement segments and becomes a nonlinear stress relaxation process. In this section, we are only interested in chain retraction as a process for stretch relaxation, and we will discuss the nonlinear aspects in section 2.5.

To generalize equations (31) - (33) to polydisperse systems, we need only update the boundary conditions:

$$\lambda(t, s = \{0, z\}, z) = 1 \qquad (34)$$

Note that the stretch relaxation equation (31) with its boundary conditions and initial conditions can be mapped to the tube survival equation for reptation (3). A distinction, however, is that the stretch relaxation equation does not contain a $z$-dependent diffusion coefficient – the length of a chain only appears in the boundary condition.

If the retracting chains are also undergoing reversible scission reactions, then we can shuffle stretch through the molecular weight distribution in the same way that was done for surviving tube segments in equation (10). Simplifying to the relevant fast breaking CLF limit, $\zeta_R \ll 1$, we once again find that to leading order all segments within a chain's interior are (on average) equally stretched everywhere, $\lambda(s, z) = \bar{\lambda}$.

At this point, we note that the parameter $\zeta_R$ also appeared in our discussion of stress relaxation by reptation and CLF in the fast breaking limit $\zeta \ll 1$. In section 2.2, we showed that for the fast breaking reptation limit $\zeta_R \gg 1$ a system can have a single relaxation time for orientation, and now we see that such a system will still possess a spectrum of relaxation times for stretch relaxation. By contrast, in the fast breaking CLF limit $\zeta_R \ll 1$ we showed that a system can have a single relaxation time for orientation and, as we now see, for stretch also (though this is not to say that they will be the same relaxation time). Once again, the tube-based depiction of stretch relaxation also requires $\zeta_R \bar{Z}^2 > 1$.

Rescaling $s$ to consider stretch relaxation very near to the chain ends, we find ($\tilde{s} = s/\zeta_R^{1/2}$):

$$0 = \frac{\partial^2 \lambda}{\partial \tilde{s}^2} - (\lambda - \bar{\lambda})\left[\int_z^\infty e^{z-z'}\lambda(\tilde{s}, z')dz' - \lambda\right] + \left[\int_0^z \lambda(s, z')dz' - z\lambda\right] \qquad (35)$$

The population balance terms here directly parallel those given in equation (21). Note that the rescaling of $s$ suggests that chain ends are able to relax their stretch for a fraction of the chain $\sim \zeta_R^{1/2}$ close to the ends. For well entangled chains with $\zeta_R \ll 1$ and $\zeta_R \bar{Z}^2 \gg 1$, this is a much larger fraction than the fraction $\sim \bar{Z}^{-1/2}\zeta_R^{1/4}$ of entanglement segments that are able to relax their orientation by CLF. This fact will become important in section 2.5 when we consider how chain retraction affects the alignment of tube segments.

The reader may also note that equation (35) assumes an equilibrium molecular weight distribution, in spite of the fact that stretched chains are more apt to break apart, which will decrease the size of a typical chain. Decoupling the reaction kinetics from the state of stress is a valid assumption provided the stretching energy held by an entanglement segment is small compared to both (1) the scission energy and (2) the activation energy for scission. This assumption breaks down under very strong flow conditions, but if the scission energy is on the order of $10 k_B T$ then there is still a wide range of flow conditions where chains are stretched but the molecular weight distribution is only weakly perturbed from its equilibrium value. The effects of flow-induced scission are considered explicitly in a forthcoming manuscript, which is partially available to view as a thesis chapter [11].

As an ansatz, suppose the solution for $\lambda(\tilde{s}, z)$ in equation (35) is independent of $z$. If this ansatz holds, then the bracketed groups in (35) are identically zero. Since $z$ does not appear in any of the remaining terms, the ansatz is affirmed and the solution to (35) is given by:

$$\lambda(\tilde{s}) = \bar{\lambda} - (\bar{\lambda} - 1)e^{-\tilde{s}} \qquad (36)$$

With time scaled by $\tau_R$, the stretch within the chain interior evolves by:

$$\partial_t \bar{\lambda} = \int_0^\infty dz\, e^{-z}\left[-2\zeta_R^{-1/2}\bar{\lambda}\right] = -2\zeta_R^{-1/2}(\bar{\lambda} - 1) \qquad (37)$$

Thus, we see that the effective stretch relation time $\tau_s$ (in units of $\tau_R$) is given exactly by $\tau_s = \zeta_R^{1/2}/2$. If the chain's orientation is also relaxing by the fast breaking CLF mechanism, then assuming a reptation time $\tau_{rep} \sim \bar{Z}^3$ and Rouse time $\tau_R \sim \bar{Z}^2$, the effective stretch relaxation time will be smaller than the effective orientation relaxation time by a factor of $\mathcal{O}\left(\bar{Z}^{1/2}\zeta_R^{1/4}\right)$. Thus, for $\zeta_R \ll 1$ and $\zeta_R \bar{Z}^2 > 1$, the effective stretch relaxation time is always faster than the timescale for relaxing orientation by CLF, but the separation between the two timescales gets smaller as the system moves deeper into the fast breaking CLF limit, $\zeta_R \ll 1$ with $\zeta_R \bar{Z}^2 > 1$.

In preparation for looking at how chain retraction affects the alignment of tube segments in section 2.5, we can use equation (36) to determine the velocity with which a chain retracts past its confining tube:

$$u(\bar{s}) = \frac{1}{\tau_R} \frac{\partial \lambda}{\partial s} = \frac{1}{2} \frac{1}{\tau_s} (\bar{\lambda} - 1) e^{-\bar{s}} \qquad (38)$$

In preparation for describing the effects of convective constraint release in section 2.7, we compute the rate at which constraints are removed by chain retraction, $f_{CCR}$. In the fast breaking CLF limit, the rate of constraint release can be computed as the mean rate of stretch relaxation scaled by the mean stretch [15]:

$$f_{CCR} = \frac{1}{\bar{\lambda}} \int_0^\infty dz \; n(z) \int_0^z ds \left( -\frac{\partial \lambda}{\partial t} \right) = \frac{2u(0)}{\bar{\lambda}} = \frac{1}{\tau_s} \left( 1 - \frac{1}{\bar{\lambda}} \right) \qquad (39)$$

To consider a full picture of chain retraction in well entangled polymers, we must also describe how retraction causes entanglement segments to become more aligned. To introduce the mathematical machinery for describing alignment (namely the tangent correlation tensor), we will first consider stress relaxation in living polymer systems of unentangled Gaussian chains relaxing by Rouse-like motion.

## 2.4. Unentangled Gaussian chains (3D Rouse)

Here, we consider stress relaxation in unentangled living polymers relaxing by Rouse-like motion. This section has immediate relevance to living polymers below the entanglement weight, but is incidentally relevant to polymers above the entanglement weight provided the breaking time is sufficiently fast, $\zeta_R \bar{Z}^2 < 1$, and chains break before entanglements can develop. Finally, this section is useful for understanding constraint release processes (Rouse-like motions of the tube in equation (80)) in fast breaking well-entangled living polymers.

To model stress relaxation in an ensemble of unentangled Gaussian chains with contours traced out by $\boldsymbol{r}(s)$, one must consider the full tangent correlation tensor describing the dyadic product of vectors tangent to the chain at contour positions $s, s' \in [0,1]$:

$$\boldsymbol{f}(s, s') = \left\langle \frac{\partial \boldsymbol{r}}{\partial s} \frac{\partial \boldsymbol{r}}{\partial s'} \right\rangle \qquad (40)$$

The tangent correlation tensor $\boldsymbol{f}(s, s')$ contains information about the typical stretch, alignment, and orientation of the Gaussian chains over the entirety of their contour lengths. For our work here, we make a distinction between orientation and alignment; orientation pertains to the second moment of a tangent vector at a particular contour position, and alignment pertains to the dyadic product of tangent vectors for two different contour positions on the same chain. Information about the orientation and alignment are found in the tangent correlation tensor $\boldsymbol{f}(s, s')$ evaluated at its diagonal, $s = s'$, and off-diagonal, $s \neq s'$, entries, respectively.

The continuum approximation of Rouse-like motion approximates finite differences in the force balance by a continuum diffusion operator. This approximation simplifies the analysis considerably, and is well-suited to describing stress relaxation processes on timescales much longer than the fastest Rouse time [24]. However, it is not well-suited to describe stress relaxation on shorter timescales where the chain can no longer be suitably described by continuum equations. With these limitations in mind, we will employ the continuum approximation of Rouse-like motion for our analysis here. In the continuum limit of a Gaussian chain, the equilibrium tangent correlation tensor tends towards a delta function at $s = s'$:

$$\boldsymbol{f}^{eq}(s, s') = \delta(s - s')\boldsymbol{I} \qquad (41)$$

The stress is given by the mean of $\boldsymbol{f}$ along $s = s'$, less the equilibrium contribution [24]:

$$\boldsymbol{\sigma} = G \int_0^1 ds \left( \boldsymbol{f}(s, s) - \boldsymbol{f}^{eq}(s, s) \right) \qquad (42)$$

In the presence of thermal fluctuations, the tangent correlation tensor evolves towards its equilibrium value by [24]:

$$\frac{\partial \boldsymbol{f}}{\partial t} = \frac{1}{\tau_R} \left( \frac{\partial^2}{\partial s^2} + \frac{\partial^2}{\partial s'^2} \right) (\boldsymbol{f} - \boldsymbol{f}^{eq}) \qquad (43)$$

where $\tau_R$ is the longest Rouse relaxation time for the Gaussian chain. Finally, the boundary conditions on the tangent correlation tensor corresponding to stress free chain ends are:

$$\boldsymbol{f}(s, 0) = \boldsymbol{f}_{eq}(s, 0) \qquad \boldsymbol{f}(s, 1) = \boldsymbol{f}_{eq}(s, 1) \qquad (44)$$

$$\boldsymbol{f}(0, s') = \boldsymbol{f}_{eq}(0, s') \qquad \boldsymbol{f}(1, s') = \boldsymbol{f}_{eq}(1, s') \qquad (45)$$

If one considers a polydisperse ensemble of chains with lengths $z$ and number density $n(z)$ then we generalize the boundary conditions to:

$$\boldsymbol{f}(s, 0, z) = \boldsymbol{f}_{eq}(s, 0) \qquad \boldsymbol{f}(s, 1, z) = \boldsymbol{f}_{eq}(s, z) \qquad (46)$$

$$\boldsymbol{f}(0, s', z) = \boldsymbol{f}_{eq}(0, s') \qquad \boldsymbol{f}(1, s', z) = \boldsymbol{f}_{eq}(z, s') \qquad (47)$$

For living polymers confined to fixed tubes, equations (8) and (9) describe how reactions shuffle information through the molecular weight distribution. Here, we propose an analogous set of rules for unentangled polymers undergoing reversible scission. When a chain of length $z_1$ combines with a chain of length $z_2$, if the original tangent correlation tensors are $\boldsymbol{f}_1$ and $\boldsymbol{f}_2$, respectively, the composite chain of length $z_3 = z_1 + z_2$ will have a tangent correlation tensor $\boldsymbol{f}_3$ given by:

$$\boldsymbol{f}_3(s, s') = \begin{cases} \boldsymbol{f}_1(s, s') & \text{if } s, s' < z_1 \\ \boldsymbol{f}_2(s - z_1, s' - z_1) & \text{if } s, s' > z_1 \\ \boldsymbol{f}^{eq}(s, s') & \text{otherwise} \end{cases} \qquad (48)$$

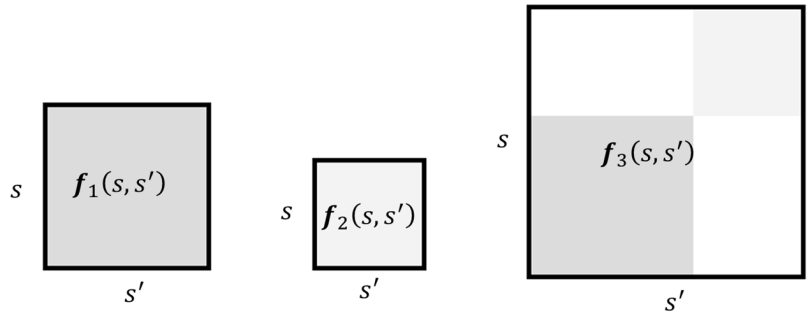

Essentially, we suggest that correlations between two chain segments on the same chain should not be affected by subsequent attachments of additional chain segments. Likewise, any two segments that were not previously part of the same chain should be (on average) uncorrelated with one another. This approximation holds provided chain ends are isotropically oriented and is therefore applicable to flexible Gaussian chains by definition but not necessarily applicable for a rod-like molecule if the monomeric rods prefer to attach only when aligned. Some additional discussion on these points is included in the third section of the appendix for interested readers.

In the reverse process, when a chain of length $z_3$ with tangent correlation tensor $\boldsymbol{f}_3$ breaks at contour position $s = z_1$, the tangent correlation tensors, $\boldsymbol{f}_1$ and $\boldsymbol{f}_2$, of the fragment chains are given by:

$$\boldsymbol{f}_1(s,s') = \boldsymbol{f}_3(s,s') \qquad \boldsymbol{f}_2(s,s') = \boldsymbol{f}_3(s+z_1, s'+z_1) \qquad (49)$$

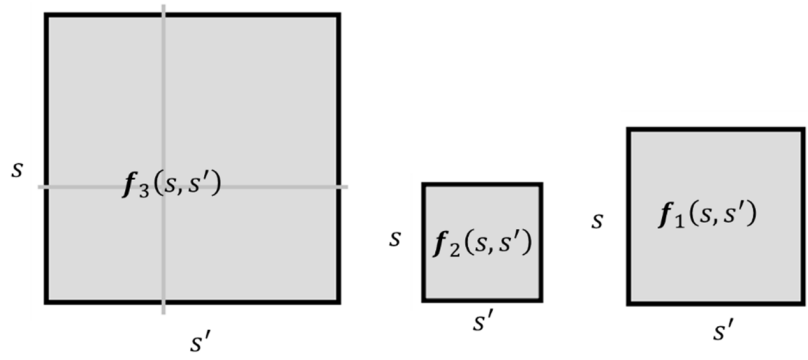

*Figure 7: Cartoon illustrating the construction of tangent correlation tensors via chain attachment, per equation* (49)

Once again, correlations between two chain segments are preserved if those chain segments are still part of the same chain after scission has occurred. Note that if the two fragments from a recently broken chain re-attach themselves, our rules in equations (48) and (49) will incorrectly show that all previously existing correlations between the two fragment chains have been completely erased. This is clearly not physical, and so the predictions of this section rely on the assumption that recently formed chain ends are more likely to find new partners than to recombine with one another[4].

Using the rules from equations (48) and (49), we can extend equation (43) to unentangled Gaussian chains undergoing reversible scission reactions. Scaling time by $\tau_R$ and assuming an equilibrium $n(z)$, we write:

$$\zeta_R \frac{\partial}{\partial t} \boldsymbol{f}(s,s',z) = \zeta_R \left( \frac{\partial^2}{\partial s^2} + \frac{\partial^2}{\partial s'^2} \right) (\boldsymbol{f} - \boldsymbol{f}^{eq}) - \boldsymbol{f} z$$

---

[4] A criterion to describe when newly-formed ends are most likely to find new partners has been previously proposed [1]. Systems that are unentangled by virtue of being dilute (as opposed to concentrated but below the entanglement weight) are more likely to self-attach, and are therefore less well suited to the modelling framework given here.

$$+ e^z \int_z^\infty dz' e^{-z'} \big( \boldsymbol{f}(s, s', z') + \boldsymbol{f}(z' - s, z' - s', z') \big)$$

$$- 2\boldsymbol{f}(s, s', z)$$

$$+ \frac{1}{2} \int_0^z dz' \left[ \begin{cases} \boldsymbol{f}(s, s', z') & \text{if} \quad s, s' < z' \\ \boldsymbol{f}(s - z', s' - z', z - z') & \text{if} \quad s, s' > z' \\ \boldsymbol{f}^{eq}(s, s') & \text{otherwise} \end{cases} \right.$$

$$+ \begin{cases} \boldsymbol{f}(s, s', z - z') & \text{if} \quad s, s' < z' \\ \boldsymbol{f}(s - z + z', s' - z + z', z') & \text{if} \quad s, s' > z' \\ \boldsymbol{f}^{eq}(s, s') & \text{otherwise} \end{cases} \Bigg] \Bigg] \qquad (50)$$

In the fast breaking limit, $\zeta_R \ll 1$, chains are rapidly broken apart and spliced together from uncorrelated fragments such that (to leading order) correlations between segments at $s$ and $s'$ are possible if and only if $s = s'$. we also see (as was true for our studies of entangled polymers) that the reactions 'shuffle' segments about, so $\boldsymbol{f}(s, s')$ must be a function of $|s - s'|$ alone. Thus, for $\zeta_R \ll 1$ the only admissible terms for the leading order solution are:

$$\boldsymbol{f}(s, s') = \boldsymbol{f}_{eq}(s, s') + \boldsymbol{f}_0 \delta_{s,s'} \qquad (51)$$

The second term (involving the Kronecker delta function $\delta_{s,s'}$) describes changes in the configuration tensor from its equilibrium value at $s = s'$. However, this solution is problematic whenever $\boldsymbol{f}_0 \neq \boldsymbol{0}$ for two reasons. First, the solution is incompatible with the boundary conditions for stress free chain ends– as in section 2.1, this indicates a boundary layer solution to describe stress relaxation at chain ends. Second, this leading order solution cannot be differentiated normal to the line of $s = s'$, as required by the stress relaxation terms in equation (50) – this indicates the presence of a boundary layer solution to resolve the transition for small values of $|s - s'|$.

Regarding the first of these issues, a boundary layer analysis will show that very near to the chain ends, a fraction of the chain $\sim \zeta_R^{1/2}$ is able to relax its stress by Rouse-like motion. If this were the only mode of stress relaxation available, the entire chain would relax its stress with a single time-scale of $\tau_B / \zeta_R^{1/2} = \tau_R \zeta_R^{1/2}$. This stress relaxation pathway has been previously reported [1].

Regarding the second issue, reactions are continually breaking chains apart and re-assembling them from uncorrelated fragments. By Rouse-like motions, those uncorrelated segments become correlated to one another and relax their stress in the process. By matching the breaking time and the Rouse time for these uncorrelated fragments, we find each uncorrelated filament covers a fraction $\zeta_R^{1/3}$ of the chain, hence there will be a boundary layer around $|s - s'| \sim \mathcal{O}(\zeta_R^{1/3})$ needed to resolve this mode of stress relaxation. The time for two adjoining but initially uncorrelated filaments of length $\zeta_R^{1/3}$ to become correlated (and relax their stress in the process) is $\tau_R \zeta_R^{2/3}$. Figure 8 provides a cartoon to illustrate the process of uncorrelated fragments becoming aligned and relaxing stress in the process.

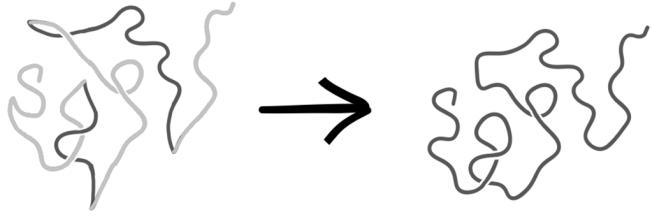

*Figure 8: In the fast breaking limit, unentangled flexible polymers rapidly break apart and re-assemble into strands of uncorrelated fragments. The 'kinks' in the chain allow stress to relax by a much faster sub-set of Rouse modes than would otherwise occur in the absence of reversible scission. The left hand side of this figure shows a polymer assembled from uncorrelated fragments (alternating grey and black) relaxing its configuration towards a more correlated state.*

For $\zeta_R \ll 1$, it is the Rouse relaxation of contiguous but initially uncorrelated fragments (rather than the Rouse relaxation of newly formed end segments) that produces the fastest stress relaxation time. Therefore, we will neglect the boundary layer for resolving stress free chain ends and only concern ourselves with understanding the transition that develops as neighboring chain segments transition from correlated to uncorrelated over contour separations of $|s - s'| \sim \zeta_R^{1/3}$.

To that end, we re-cast equation (50) from its original coordinate system in $s$ and $s'$ to a coordinate system rotated by 45 degrees with axes defined by the lines $s = s'$ and $s = 1 - s'$:

$$\zeta_R \frac{\partial}{\partial t} \boldsymbol{f}(p, p', z) = \frac{\zeta_R}{2} \left( \frac{\partial^2}{\partial p^2} + \frac{\partial^2}{\partial p'^2} \right) \boldsymbol{f} + \cdots \qquad (52)$$

$$p = \frac{1}{2}(s - s') \qquad p' = \frac{1}{2}(s + s') \qquad (53)$$

We will need to rescale the $p$-axis to resolve the leading order boundary layer solution, but if a suitable rescaling exists, then we note that the ensuing solution will show no variation in $p'$ or $z$ (due to the 'shuffling' effect of the reaction terms). Therefore, before rescaling in $p$, we integrate the reaction terms in $p'$ and $z$ while assuming $f(p, p', z) = f(p)$. In doing so, we find that the population balance terms in equation (50) reduce to:

$$\frac{\partial}{\partial t} \boldsymbol{f} = \cdots - \frac{2|p|}{\zeta_R} (\boldsymbol{f} - \boldsymbol{f}^{eq}) \qquad (54)$$

What survives from the population balance terms can be interpreted as follows: information about correlations between points at $s_1$ and $s_2$ is lost whenever a break occurs at a point $s \in [s_1, s_2]$. The rate at which a break occurs in this range is proportional to the length of the chain between the two points, $|s_2 - s_1| = 2|p|$ and inversely proportional to the dimensionless breaking time, $\zeta_R$.

Now rescaling in $p$ and $t$ we obtain a leading order boundary layer equation ($\tilde{p} = p/\zeta_R^{1/3}$, $\tilde{t} = t/\zeta_R^{2/3}$):

$$\frac{\partial \boldsymbol{f}}{\partial \tilde{t}} = \left( \frac{1}{2} \frac{\partial^2}{\partial \tilde{p}^2} - 2|\tilde{p}| \right) (\boldsymbol{f} - \boldsymbol{f}^{eq}) \qquad (55)$$

with boundary conditions:

$$\boldsymbol{f}(\tilde{t}, \tilde{p} \to \pm\infty) = 0 \qquad (56)$$

If the initial condition is chosen to reflect a small step strain (shear) of amplitude $\gamma$ imposed on an initially equilibrated system (as needed for studying linear viscoelasticity) then we also need to incorporate the effects of deformation and flow into equation (55). Since we have assumed our chains are flexible, their deformation is affine and can be represented via an upper convected Maxwell derivative in (54). Thus, we obtain our initial condition to a small amplitude initial step strain, $\gamma \ll 1$:

$$\boldsymbol{f}(0, \tilde{p}) = \boldsymbol{f}^{eq}(\tilde{p}) + \gamma f_{yy}^{eq}(\tilde{p})\big(\boldsymbol{e}_x \boldsymbol{e}_y + \boldsymbol{e}_y \boldsymbol{e}_x\big) + \mathcal{O}(\gamma^2) \qquad (57)$$

Constitutive modelling for the linear rheology of fast breaking unentangled Gaussian chains is not a primary focus of this paper, but having developed the equations we will dedicate the remainder of this subsection to evaluating and interpreting their predictions. To evaluate, we first eliminate the delta function: after a Laplace transform, the delta function can be absorbed into the boundary conditions on $\tilde{p}$:

$$\boldsymbol{g}(\tilde{p}, q) = \frac{1}{\gamma} \int_0^\infty e^{-qt} \big( \boldsymbol{f}(\tilde{t}, \tilde{p}) - \boldsymbol{f}^{eq}(\tilde{p}) \big) d\tilde{t} \qquad (58)$$

$$q g_{xy} = \left( \frac{1}{2} \frac{\partial^2}{\partial \tilde{p}^2} - 2|\tilde{p}| \right) g_{xy} \qquad (59)$$

$$\frac{\partial g_{xy}}{\partial \tilde{p}} (\tilde{p} = 0^+) = -1 \qquad g_{xy}(\tilde{p} \to \infty) = 0 \qquad (60)$$

This admits an analytic solution:

$$g_{xy}(\tilde{p}, q) = -\frac{1}{2^{2/3}} \frac{\text{Ai}\left( \frac{q + 2|p|}{2^{1/3}} \right)}{\text{Ai}'\left( \frac{q}{2^{1/3}} \right)} \qquad (61)$$

where $\text{Ai}(x)$ is the Airy function and $Ai'(x)$ is its derivative. From this solution, one can obtain the complex modulus, $G^*(\omega)/G_e = i\omega g_{xy}(0, i\omega)$, where $G_e$ is the a shear modulus for a single Rouse mode and $\omega$ is a dimensionless frequency (in units of $1/\tau_R \zeta_R^{2/3}$) for an imposed oscillatory small amplitude deformation. The loss modulus, $G''(\omega) = \mathcal{I}\big(G^*(\omega)\big)$ and storage modulus $G'(\omega) = \mathcal{R}\big(G^*(\omega)\big)$ for a fast breaking Gaussian chain are shown in Figure 9. The rheological predictions are reminiscent of (but not equivalent to) the predictions that one obtains for monodisperse, unbreaking, unentangled Gaussian chains.

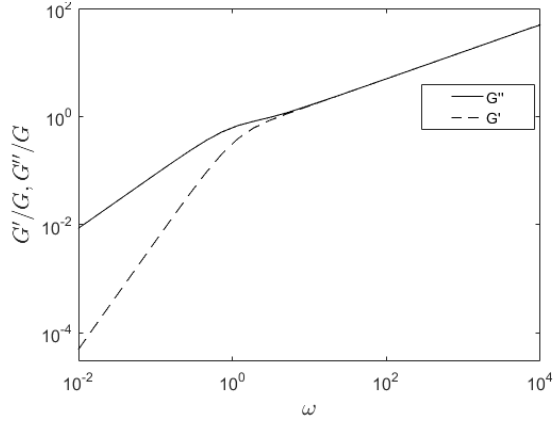

*Figure 9: Predictions for the linear rheology, $G'$ and $G''$ of unentangled fast breaking living polymers, $\zeta_R \ll 1$ (or $\zeta_R \bar{Z}^2 \ll 1$ for systems above the entanglement weight). The predictions are similar to predictions that one would obtain for monodisperse unbreaking polymers with the same terminal relaxation time.*

Whereas all the previous 'fast breaking' limits produced a single relaxation time-scale, the fast breaking limit of an unentangled Gaussian chain produces a very Rouse-like spectrum of relaxation times. The single relaxation time observed for entangled chains arises because scission creates new opportunities for relaxation of chain ends – for unentangled chains, however, the speedup of stress relaxation is more directly attributed to recombination (not scission) moving stress into a fast-relaxing subset of the available Rouse modes. Thus, for fast breaking unentangled chains, we essentially see the whole Rouse relaxation spectrum for chain-scales below the typical uncorrelated filament size.

Finally, we note that the analysis in this section assumed that chain ends never recombine with their recent neighbors. Practically speaking, many systems of unentangled living polymers are unentangled by virtue of being dilute, in which case self-healing of chain ends may be common [1]. If the probability for a pair of chain ends to self-heal is given by $P_r$, then for small values of $P_r$ we can suppose that the primary effect of self-healing is to simply re-scale the breaking time, where only breaks that do not self-heal are considered to be 'true' breaks. This suggests a revision to equation (55):

$$\frac{\partial \boldsymbol{f}}{\partial t} = \left(\frac{1}{2}\frac{\partial^2}{\partial \bar{p}^2} - 2(1 - P_r)|\bar{p}|\right)(\boldsymbol{f} - \boldsymbol{f}^{eq}) \qquad (62)$$

Similar tricks might be applied whenever recombination is biased to favor ends that are aligned with one another, as we will discuss in section 2.7.

Having introduced the concept of a tangent correlation tensor for describing alignment of tube segments, we can return to our discussion of stretch relaxation in the tube to show how retraction affects the alignment and orientation of tube segments in the fast breaking CLF limit.

## 2.5. Chain retraction (Stretch, orientation, and alignment)

When a well entangled polymer relaxes its stretch by chain retraction in a fixed confining tube, the number of tube segments that separate any two points along the chain's contour decreases, thereby increasing (on average) the degree to which entanglement segments separated by a fixed chain interval are aligned with one another.

In this section, we will momentarily ignore reactions except to the extent that we assume a pre-existing state of stretch given by the analysis of section 2.3. Thereafter, we only consider stress relaxation dynamics for a single unbroken chain.

During chain retraction, the contour of a chain evolves by:

$$\frac{\partial \boldsymbol{r}}{\partial t} = \hat{u}(s)\frac{\partial \boldsymbol{r}}{\partial s} \qquad (63)$$

where $\hat{u}(s) = u(s)/\lambda(s)$ is the velocity with which a chain slides past its confining tube, scaled by the local state of stretch. This can be obtained from the 1D chain retraction problem (c.f. section 2.3), and for the analysis given here we assume that the fast breaking CLF limit applies, $\zeta_R \ll 1$. Given equation (63), the tangent correlation tensor evolves by:

$$\frac{\partial \boldsymbol{f}}{\partial t} = \left[\hat{u}'(s) + \hat{u}'(s') + \hat{u}(s)\frac{\partial}{\partial s} + \hat{u}(s')\frac{\partial}{\partial s'}\right]\boldsymbol{f} \qquad (64)$$

Rotating our coordinate system as in equation (52), this is equivalent to:

$$\frac{\partial \boldsymbol{f}}{\partial t} = \left[\hat{u}'(p'-p) + \hat{u}'(p'+p) + \frac{1}{2}\hat{u}(p'-p)\left(\frac{\partial}{\partial p} + \frac{\partial}{\partial p'}\right) + \frac{1}{2}\hat{u}(p'+p)\left(\frac{\partial}{\partial p'} - \frac{\partial}{\partial p}\right)\right]\boldsymbol{f} \qquad (65)$$

If we assume that chain retraction is occurring in a system of well entangled living polymers with $\zeta_R \ll 1$ and $\zeta_R \bar{Z}^2 > 1$ (relaxing orientation by CLF and stretch by chain retraction) then the tangent correlation tensor will not vary in $p'$ except when chain ends appear, i.e. near the boundaries in $p, p'$. We also note that the boundary layer for orientation is small, of order $\zeta^{1/4}\bar{Z}^{-1/4}$, compared to the boundary layer for stretch relaxation, of order $\zeta_R^{1/2}$. Thus, when considering the effects of chain retraction one can safely ignore the boundary layer created by CLF.

Next, in the fast breaking CLF limit, relaxation of alignment between entanglement segments is primarily achieved by scission and recombination unless the two entanglement segments are quite close together along the chain, $|p| \ll 1$. Therefore we can linearize equation (65) for small $|p|$ to yield an expression that is much simpler but still preserves the same essential behavior for the range of $p$ where retraction is most important.

$$\frac{\partial \boldsymbol{f}}{\partial t} \approx \hat{u}'(p')\left[2 + p\frac{\partial}{\partial p}\right]\boldsymbol{f} \qquad (66)$$

With a few additional approximations, we show next that the remaining dependence on $p'$ can be integrated out. When the chain is not strongly stretched, $\bar{\lambda} - 1 \ll 1$, the tangent correlation tensor $\boldsymbol{f}$ primarily contains information about the alignment and orientation of chains. Relaxation of orientation and stretch both occur only very near to the chain ends, but recall that stretch relaxation penetrates farther into the chain interior. As a first approximation, we can assume that the portion of the tangent correlation tensor describing orientation and alignment is constant across the portion of chain undergoing retraction. With this assumption, the tangent correlation tensor $\boldsymbol{f}(p', p)$ can be related to the stretch $\lambda(p')$ and orientation/alignment tensor $\boldsymbol{S}$. We define the orientation tensor as:

$$\boldsymbol{S}(s, s') = \langle\frac{1}{\lambda(s)}\frac{\partial \boldsymbol{r}}{\partial s}\frac{1}{\lambda(s')}\frac{\partial \boldsymbol{r}}{\partial s'}\rangle \qquad (67)$$

Since stretch relaxation penetrates deeper into a chain than relaxation of orientation, for most of the region where stretch relaxation occurs we can approximate:

$$\boldsymbol{S}(s, s') \approx \boldsymbol{S}(s - s') = \boldsymbol{S}(p) \qquad (68)$$

Finally, for $p \ll 1$ we linearize to obtain:

$$\boldsymbol{f}(p, p') \approx \lambda(p' - p)\lambda(p' + p)\boldsymbol{S}(p) \approx \lambda(p')^2 \boldsymbol{S}(p) \qquad (69)$$

Using (69), we can integrate equation (66) over $p'$ and over the entire molecular weight distribution to obtain:

$$\frac{\partial}{\partial t}\boldsymbol{f}(t, p) = -\frac{1}{\tau_s}\left(1 - \frac{1}{\bar{\lambda}}\right)\left(2 + p\frac{\partial}{\partial p}\right)\boldsymbol{f}(t, p) \qquad (70)$$

In equation (70), the prefactor $(1 - 1/\bar{\lambda})/\tau_s$ effectively gives the rate of chain retraction. The second grouping of terms, $2 + p\partial_p$, describes (via the first term) the change in $\boldsymbol{f}$ arising due to relaxation of stretch alone, and (via the the second) how retraction increases alignment. Regarding the latter, as chains retract within a fixed tube, the alignment of any two tube segments remains fixed but the number of tube segments separating any two chain segments decreases. Thus, the motion of retraction causes chain segments to become more aligned with neighboring segments, with the effect being larger for increasing separation between segments.

In the fast breaking CLF limit, chain retraction does not affect the orientation of tube segments, since all tube segments (except for those within the narrower CLF boundary layer) are sampled from the same orientation distribution. This is in contrast to unbreaking polymers, where retraction causes the most isotropic entanglement segments (nearer to the chain ends) to retract into tube segments that are more likely to be oriented in a particular direction.

At this stage, the mean stretch $\bar{\lambda}$ can be estimated directly from the $p'$-averaged tangent correlation tensor via the closure approximation:

$$\bar{\lambda} = \left\langle\left|\frac{\partial r}{\partial s}\right|\right\rangle \approx \left[\left\langle\left|\frac{\partial r}{\partial s}\right|^2\right\rangle^{1/2}\right] = \left[\frac{1}{3}\operatorname{tr}(\boldsymbol{f}(p = 0))\right]^{1/2} \qquad (71)$$

In the fast breaking CLF limit, this closure approximation is especially appropriate: stretch relaxes at chain ends, but then the relaxed segments attach to one another and are re-incorporated into a chain interior. Long before another break occurs in the vicinity of the recent attachment, the stretch relaxed near the break point will equilibrate with its surrounding chain segments. Thus, the stretch is homogenized in interior chain segments not only in an ensemble-averaged sense but also for individual chains. With increasingly homogenized stretch, the closure approximation (71) becomes exact.

It is worth noting that equations (70) and (71) are also consistent with the 1D picture of chain retraction (c.f. section 2.3), where the mean stretch relaxed towards its equilibrium value $\bar{\lambda} = 1$ with a single time-scale $\tau_S$:

$$\frac{\partial}{\partial t}\left[\frac{1}{3}\operatorname{tr}(\boldsymbol{f}(p = 0))\right]^{\frac{1}{2}} = \frac{1}{2\bar{\lambda}}\frac{\partial}{\partial t}\bar{\lambda}^2 = \frac{\partial\bar{\lambda}}{\partial t} = \frac{1}{2\bar{\lambda}}\left(-\frac{2}{\tau_s}\left(1 - \frac{1}{\bar{\lambda}}\right)\bar{\lambda}^2\right) = -\frac{1}{\tau_s}(\bar{\lambda} - 1)$$

Finally, we would like to note that a similar depiction of chain retraction in wormlike micelles has previously been given by Milner et. al. [16]. A key difference, however, is that our picture of chain retraction allows for a finite stretch relaxation time and is therefore more useful for modelling real systems with finite entanglement numbers.

This concludes our discussion of the various modes of stress relaxation available to well entangled living linear polymers. In section 2.7, we will combine everything into a single framework to make predictions for the linear and nonlinear rheology of well entangled living polymers in the fast breaking CLF limit. But first, we will take a moment to review all of the main results to this point.

## 2.6. Summary of results: terminal relaxation times

In this section, we will summarize results of the preceding sections for the terminal relaxation of orientation, $\tau$, and stretch, $\tau_S$, in living polymers for which reversible scission and/or end attack are the dominant reaction pathways relevant to stress relaxation dynamics.

For $\zeta \gg 1$, $\tau \sim \tau_{rep}$ and $\tau_S \sim \tau_R$ with $\tau/\tau_S \sim \bar{Z}$. Both orientation and stretch relax exhibit a broad spectrum of relaxation times.

When chains are fast breaking with respect to reptation, $\zeta \ll 1$, the details of stress relaxation dynamics fall into three sub-categories depending on the value of $\zeta_R$. When $\zeta_R > 1$, we call this the 'fast breaking reptation limit' since stress relaxation is dominated by reptation at newly-formed chain ends. When $\bar{Z}^{-2} < \zeta_R < 1$, we call this the 'fast breaking CLF limit' since stress relaxation is dominated by CLF at newly-formed chain ends. When $\zeta_R < 1/\bar{Z}^2$, we call this the 'post-entangled limit', since chains break apart too quickly to preserve any meaningful concept of entanglement . The terminal relaxation times for stretch and orientation in each of these regions are as follows:

In the fast breaking reptation limit, we find $\tau \sim \tau_{rep}\zeta^{1/2}$ and $\tau_S \sim \tau_R$ with $\tau/\tau_S \sim \bar{Z}\zeta^{1/2}$. Excluding the effects of constraint release, orientation relaxes with a single timescale but stretch does not. The fast breaking reptation limit was the focus of section 2.1.

In the fast breaking CLF limit, we find $\tau \sim \tau_{rep}\zeta^{1/2}\zeta_R^{1/4}$ and $\tau_S \sim \tau_R\zeta_R^{1/2}$ with $\tau/\tau_S \sim \bar{Z}^{1/2}\zeta_R^{1/4}$. Excluding the effects of constraint release, both orientation and stretch each relax on a single (distinct) timescale. The fast breaking CLF limit was the focus of sections 2.2 and 2.3.

In the post-entangled limit, we find $\tau = \tau_S \sim \tau_R\zeta_R^{1/2}$. Neither orientation nor stretch relax on a single timescale. The post-entangled limit is an incidental subject of section 2.4.

These scaling results for linear relaxation times are mostly known in the existing literature, but were previously derived using methods that cannot be extended to model non-linear rheology [1] [10] [3]. The scaling we report for the post-entangled limit differs from the most highly cited result [1] but has previously been derived for fast breaking living polymers [9].

Having completed our summary of the preceding sections, we will proceed to develop a full nonlinear constitutive model for well entangled living linear chain polymers in the fast breaking CLF limit. The fast breaking CLF limit is a natural starting point for nonlinear constitutive modelling of living polymers, since it is the only one for which both stretch and orientation relax on a single timescale.

Nonlinear constitutive models for slower-breaking systems will be the subject of a follow-up report, and section 2.4 includes a constitutive model for the post-entangled limit (use an upper convected Maxwell derivative in equation (55)).

## 2.7. Putting it all together

To construct a full nonlinear rheology model of well-entangled living polymers, one could in principle repeat the analysis of the preceding sections beginning with something like the GLaMM model [15] to describe stress relaxation in the absence of reactions. The resulting model would look similar to equation (50), but with the relaxation dynamics and equilibrium tangent correlation tensor adapted for the unbreaking polymer model of interest. However, the constitutive model produced in this way is too complex to be of any practical use unless it is first simplified. We do this here with the fast breaking CLF limit in mind. The details of such a simplification are tedious but ultimately equivalent to what has already been covered in the preceding sections. Therefore, for simplicity, we present the final result that emerges from such an analysis as a piece-wise assembly of our earlier results. The model given here is suitable for well entangled living polymer systems in the fast breaking CLF limit $\zeta_R \ll 1$ and $\zeta_R \bar{Z}^2 > 1$.

Considering the effects of deformation, we assume that chains are flexible and can be deformed affinely in flow. We do not allow for any 'slipping' against the background deformation, since the molecular relaxation processes associated with slipping (e.g. chain retraction) are explicitly encoded in subsequent relaxation dynamics. Given a flow with velocity field $\boldsymbol{u}$, the contour of the confining tube $\boldsymbol{r}(s)$ as a function of its arc-length evolves by:

$$\partial_t \boldsymbol{r}(s) = \boldsymbol{r} \cdot \nabla \boldsymbol{u} + \cdots$$

and the tangent correlation tensor $\boldsymbol{f}(s, s') = \langle \boldsymbol{r}'(s)\boldsymbol{r}'(s') \rangle$ then evolves by:

$$\frac{\partial}{\partial t} \boldsymbol{f}(t, \boldsymbol{p}) = -\boldsymbol{u} \cdot \nabla \boldsymbol{f} + (\nabla \boldsymbol{u})^T \cdot \boldsymbol{f} + \boldsymbol{f} \cdot \nabla \boldsymbol{u} + \cdots \qquad (73)$$

For convenience, we can also write this as the upper convected Maxwell derivative:

$$\overset{\triangledown}{\boldsymbol{f}} = \frac{\partial \boldsymbol{f}}{\partial t} + \boldsymbol{u} \cdot \nabla \boldsymbol{f} - (\nabla \boldsymbol{u})^T \cdot \boldsymbol{f} - \boldsymbol{f} \cdot \nabla \boldsymbol{u} \qquad (74)$$

Chains relax orientation and alignment by CLF with a single relaxation time, $\tau = 0.26 \tau_{rep} \zeta^{1/2} \zeta_R^{1/4}$, per section 2.2 and equation (30). Section 2.2 primarily discusses relaxation of orientation, but the generalization to alignment is trivial: two chain segments relax their alignment by CLF whenever one of the two segments relaxes its orientation by CLF:

$$\frac{\partial \boldsymbol{f}}{\partial t} = \cdots - \frac{1}{\tau}(\boldsymbol{f} - \boldsymbol{f}^{eq}) + \cdots \qquad (75)$$

For the equilibrium tangent correlation tensor, we describe the tube as a freely jointed chain with segment lengths sampled from a Poisson distribution. Scaling $p$ by twice of the mean entanglement length (i.e. $p = 1/2$ represents a separation equal to the mean entanglement length), we write the equilibrium tangent correlation tensor as:

$$\boldsymbol{f}_{eq}(p) = \boldsymbol{I} \exp(-4|p|) \qquad (76)$$

If ever the persistence length of a chain exceeds the typical entanglement length, then this estimate of the equilibrium tangent correlation tensor breaks down. Under such conditions, the equilibrium tangent correlation tensor will just be the equilibrium tangent correlation tensor of a semi-flexible chain:

$$\boldsymbol{f}_{eq}(p) = \boldsymbol{I} \exp(-4\alpha_e |p|) \qquad (77)$$

where $1/\alpha_e$ is the number of entanglement segments per persistence length. Chains retract within their tubes, reducing stretch and increasing alignment, with a stretch relaxation time $\tau_s = \tau_R \zeta_R^{1/2}/2$. The details of chain retraction are given in sections 2.3 and 2.5, culminating in equation (70)

$$\frac{\partial \boldsymbol{f}}{\partial t} = \cdots - \frac{1}{\tau_s}\left(1 - \frac{1}{\bar{\lambda}}\right)\left(2 + p\frac{\partial}{\partial p}\right)\boldsymbol{f} + \cdots \qquad (78)$$

Alignment of entanglement segments also relaxes through reversible scission, as discussed in section 2.4. If the breaking time for two segments separated by a distance $|p|$ is $\tau_B/2|p|$, then assuming chain ends are isotropically oriented the rate at which correlations between tube segments are lost as a direct result of reversible scission is:

$$\frac{\partial \boldsymbol{f}}{\partial t} = \cdots - \frac{2|p|}{\bar{Z}\tau_B}(\boldsymbol{f} - \boldsymbol{f}^{eq}) + \cdots \qquad (79)$$

Since the approximation of isotropically oriented ends may not always be true of semi-flexible and stiff polymers (including wormlike micelles), equation (79) should be understood as an upper-bound estimate on the rate of orientational relaxation occurring by scission directly. When chain ends are not isotropic and alignment is a prerequisite for end attachment, this term can be reduced or removed altogether, c.f. equation (62). In any case, we have found that this term has a negligible effect on the model predictions – at small $|p|$ alignment is primarily relaxed by CLF, and the alignment at large $|p|$ does not seem to affect rheological predictions in any meaningful way.

Finally, we consider stress relaxation from constraint release Rouse relaxations of the tube itself. Since the terminal constraint release Rouse time of the tube is very long compared to the breaking time, we can use the results of section 2.4 and equation (55) to describe the effect of Rouse relaxations on timescales longer than the fastest constraint release Rouse relaxation time. For timescales comparable to or faster than the fastest Rouse relaxation time, a continuum approximation of the Rouse model is not appropriate and additional work will be required.

Given a constraint release frequency $\nu$ (units of 1/time, representing the rate at which a tube segment 'hops' a distance equal to its own diameter) we write[5]:

$$\frac{\partial \boldsymbol{f}}{\partial t} = \cdots + \frac{3}{4}\nu\frac{\partial^2}{\partial p^2}(\boldsymbol{f} - \boldsymbol{f}^{eq}) \qquad (80)$$

---

[5] Some authors have suggested that chain stretching reduces the efficacy of constraint release Rouse relaxations in well-entangled linear polymer systems [15], but we omit these corrections on the grounds that they are still speculative and have not yet been independently established by simulations or experiments.

The constraint release frequency is a sum of the frequencies for constraint release by thermal motion (dominated by CLF) and by chain retraction, $\nu = \nu_{CLF} + \nu_{ret}$. Regarding the former, $\nu_{CLF}$ is inversely proportional to the mean survival time for a tube segment, which in this case is $\tau_{CLF}$. Regarding the latter, $\nu_{ret}$ is proportional to the rate at which tube segments are lost by retraction (see equation (39) in section 2.3 for reference). The constant of proportionality, $c_\nu$, is assumed to be the same for both channels of constraint release and independent of the polymer configuration:

$$\nu = c_\nu \left( \frac{1}{\tau_{CLF}} + \frac{1}{\tau_s}\left(1 - \frac{1}{\bar{\lambda}}\right) \right) \qquad (81)$$

Nominally, the constraint release parameter $c_\nu$ is inversely proportional to the number of constraint release events needed to allow a tube segment to move a distance comparable to its own diameter [15]. Therefore, $c_\nu$ is typically bounded by $c_\nu \in [0,1]$ with monodisperse unbreaking flexible linear polymers typically assigned a value around $c_\nu = 0.1$ [16] [15].

Scaling time by the orientational relaxation time $\tau$, we re-write the entire constitutive equation in dimensionless form, combining equations (73) −(81):

$$\overset{\triangledown}{\boldsymbol{f}} = \left[ -1 + c_\nu \left( 1 + \theta\left(1 - \frac{1}{\bar{\lambda}}\right) \right) \frac{\partial^2}{\partial p^2} - 0.4\frac{|p|}{\theta} \right] (\boldsymbol{f} - \boldsymbol{f}^{eq}) - \theta\left(1 - \frac{1}{\bar{\lambda}}\right)\left(2 + p\frac{\partial}{\partial p}\right)\boldsymbol{f} \qquad (82)$$

$$\bar{\lambda} = \left[ \frac{1}{3}\text{tr}(\boldsymbol{f}(p=0)) \right]^{1/2} \qquad (83)$$

$$\boldsymbol{f}^{eq} = \boldsymbol{I}\exp(-4p) \qquad (84)$$

where $\theta = \tau/\tau_s = 0.9\bar{Z}^{1/2}\zeta_R^{1/4}$ is the ratio of the relaxation times for orientation and stretch and the 0.4 prefactor on the scission term appears via the product of several previously listed $O(1)$ prefactors (e.g. the 0.26 prefactor in $\tau$). In the fast breaking CLF limit (where our model applies) $\theta$ will always be greater than one but less than $\bar{Z}^{1/2}$. Boundary conditions are given by symmetry at $p = 0$ (because of the diffusion term) and by the absence of correlations between segments infinitely far from one another:

$$\frac{\partial}{\partial p}(\boldsymbol{f} - \boldsymbol{f}^{eq})(p = 0) = \boldsymbol{0} \qquad \boldsymbol{f}(p \to \infty) = \boldsymbol{0} \qquad (85)$$

Finally, the polymer stress $\boldsymbol{\sigma}$ is related to the tangent correlation tensor by the shear modulus $G_e$:

$$\boldsymbol{\sigma} = G_e(\boldsymbol{f} - \boldsymbol{f}^{eq})_{p=0} \qquad (86)$$

For the special case of $c_\nu = 0$, the diffusion terms and the zero gradient boundary condition can be dropped. In what remains, there are no couplings between the tangent correlation tensor at different values of $p$, and the rheology (which depends on $\boldsymbol{f}(p=0)$) reduces to a single-mode model, namely the Rolie Poly model [19] with CCR turned off.

As it happens, a constitutive model very similar to the one given here was proposed nearly two decades ago by Milner, McLeish, and Likhtman (MML) [16]. Using the notation defined in this paper, the MML model can be written as:

$$\overset{\triangledown}{\boldsymbol{f}} = \left[-1 + c_\nu(1+\Lambda)\frac{\partial^2}{\partial p^2}\right](\boldsymbol{f} - \boldsymbol{f}^{eq}) - \Lambda\left(2 + p\frac{\partial}{\partial p}\right)\boldsymbol{f} \qquad (87)$$

$$\boldsymbol{f}^{eq} = \boldsymbol{I}\delta(p) \qquad (88)$$

$$\frac{\partial}{\partial p}(\boldsymbol{f} - \boldsymbol{f}^{eq})(p=0) = \boldsymbol{0} \qquad \boldsymbol{f}(p \to \infty) = \boldsymbol{0} \qquad (89)$$

where $\Lambda$ is a Lagrange multiplier that ensures no-stretching, $\text{tr}(\boldsymbol{f} - \boldsymbol{f}^{eq})_{p=0} = 0$. Compared to the STARM model, the MML model for living polymers was built on similar intuitions and has a similar mathematical form but ultimately lacked a number of important features, including: (1) an unambiguous derivation for the final simplified constitutive equation; (2) a physically meaningful finite size cutoff for the equilibrium tangent correlation tensor; (3) the ability to handle finite stretch relaxation times and; (4) fast convergence of numerical solutions. It is our view that our work formalizes and sharpens the intuitions of the MML authors and yields a more useful constitutive model in the process.

From this point forward, we will refer to equation (82) as the 'Simplified Tube Approximation for Rapid-breaking Micelles', or STARM model.

The STARM model has four (or five) adjustable parameters: the shear modulus $G_e$, the terminal relaxation time $\tau$, the ratio of relaxation times $\theta$, the constraint release parameter $c_\nu$, and (whenever $\alpha_e < 1$) the number of persistence lengths per entanglement $\alpha_e$. In practice, however, $\alpha_e$ and $c_\nu$ appear to be degenerate parameters: aligned tube segments will remain aligned if (1) chains prefer to remain aligned, $\alpha_e \ll 1$ or (2) there is no mechanism by which alignment can be relaxed, $c_\nu \ll 1$. When the term describing decorrelation via breaking (79) can be ignored (as it often can) the net effect of these two parameters depends only on the product $c_\nu^{eff} = c_\nu \alpha_e^2$. Therefore, for the present study we will only consider varying $c_\nu$ with fixed $\alpha_e = 1$ (i.e. assuming flexible entanglement segments) and the effects of varying $\alpha_e$ for stiff entanglements can be inferred therefrom.

As a linear rheology model, STARM applies equally well to the fast breaking reptation and fast breaking CLF regime, since both of these are characterized by stress relaxation concentrated at chain ends and a single overall relaxation time. As a non-linear rheology model, STARM is not ideally suited to the fast breaking reptation regime since it fails to capture the spectrum of stretch relaxation times. However, if the spectrum of stretch relaxation times can be reduced to a single mode approximation, the STARM model will be a very good small-mode approximation for the fast breaking reptation regime.

# 3. Predictions of the STARM model

## 3.1. Linear rheology of the STARM model

Here, we present a summary of results for the linear rheology of the STARM model. Linearizing equation (82) for stress relaxation of a small amplitude step strain, $\gamma \ll 1$, we find that the complex modulus $G^*(\omega) = G'(\omega) + iG''(\omega)$ satisfies:

$$\frac{G^*(\omega)}{G_e} = i\omega g_{xy}(0, i\omega) \qquad (90)$$

$$qg_{xy} - \exp(-4|p|) = \left[-1 + c_\nu\frac{\partial^2}{\partial p^2} - 0.4\frac{|p|}{\theta}\right]g_{xy} \qquad (91)$$

$$\frac{\partial g_{xy}}{\partial \tilde{p}}(\tilde{p}=0) = 0 \qquad g_{xy}(\tilde{p} \to \infty) = 0 \qquad (92)$$

These equations can be solved numerically without much difficulty. Our method of choice transforms the unbounded domain $p > 0$ to a bounded domain $x = -1 + 2p/(1+p)$ and then applies Chebyshev collocation. This procedure gives accurate results with less than one hundred grid points, so calculations are quite fast compared to other 'full chain' constitutive models that include constraint release [15] [16] [25].

Fixing $c_\nu = 0.1$, a 'standard' value for monodisperse unbreaking polymers [15] [16], Figure 10 shows that varying $\theta$ for $\theta = 1, 10, 100$ has virtually no meaningful effect on the linear rheology. Predictions of linear rheology are practically independent of $\theta$ for other values of $c_\nu$ as well.

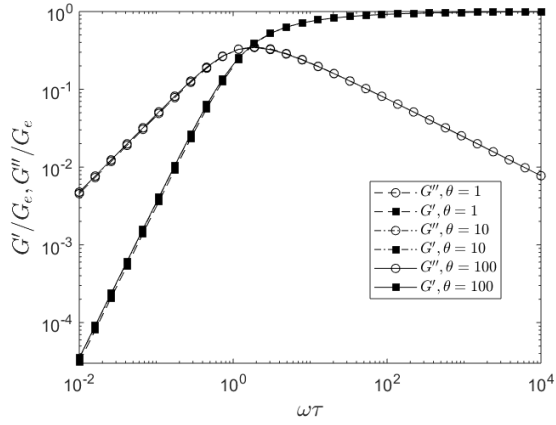

Figure 10: Predictions for the linear rheology, $G'$ (open circles) and $G''$ (filled squares) of well entangled fast breaking living polymers with $c_\nu = 0.1$ and a range of values for $\theta = 1$ (dashed) $\theta = 10$ (dash-dotted) and $\theta = 100$ (solid). The predictions are similar to what one finds for monodisperse unbreaking polymers with the same terminal relaxation time.

Recalling that $\theta > 1$ whenever our model assumptions apply, we conclude that the removal of correlations by scission (c.f. equation (79)) has very little effect on the linear rheology. For chain segments separated by $\theta$ entanglements, scission is the dominant means by which correlations are relaxed, but for chain segments spaced more closely together only reptation and constraint release are important. Thus, to determine the stress on a chain (related to $f$ evaluated at $p = 0$) we can ignore interactions of scission and constraint release.

Ignoring completely the loss of correlations by scission (i.e. assuming $\theta \gg 1$) we find that equations (90) - (92) have an analytic solution:

$$\frac{G^*(\omega)}{G_e} = \left(\frac{B}{16-A}\right)\left[1 - \frac{4}{\sqrt{A}}\right] \qquad (93)$$

$$B = -\frac{4}{3c_\nu}i\omega \qquad A = \frac{4}{3c_\nu}(1+i\omega) \qquad (94)$$

For the remainder of this section, we will present results for $\theta \gg 1$ but practically indistinguishable results can be produced for any $\theta > 1$.

The astute reader may note that linear rheology predictions shown in Figure 10 do not have the 'single Maxwell' response that one might usually expect from fast breaking living polymers [1] [20] [26]. Whereas there is a single relaxation time for a tube segment to relax its stress by reptation and breaking, the same is not true after accounting for the orientational relaxation that occurs by a continuum approximation of the tube's Rouse-like motions (c.f. Figure 9). Similar departures from the 'single Maxwell' response have also been previously predicted for systems outside of the fast breaking limit, $\zeta > 1$ [10] [3], but it is important to note that this is not the mechanism at work here. A single overall relaxation time (i.e. a pure Maxwell fluid response) is recovered in STARM model if and only if constraint released is suppressed relative to what is typical of unbreaking flexible polymers. In Figure 11, we show Cole-Cole plots for $c_v$ varying from 0 to 0.1, and the model predictions are a poor approximation of a semi-circle unless $c_v \ll 10^{-2}$.

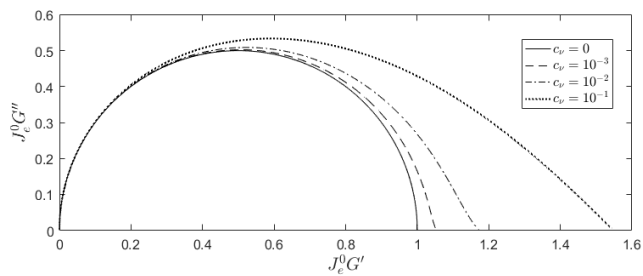

Figure 11: *Predictions for the linear rheology, $G'$ and $G''$ of well entangled fast breaking living polymers with varying $c_v$ and $\theta \gg 1$. Note that the axis are scaled by the steady shear recoverable compliance modulus, $J_e^0$, to ensure that predictions collapse for all values of $c_v$ approaching the origin.*

Thus, the STARM model predicts that a system of living polymers cannot have a Maxwell-like linear rheology unless it is both fast breaking and has negligible stress relaxation by Rouse motion of the tube $c_v \ll 10^{-2}$. From existing studies on flexible unbreaking polymers, there is a general consensus that $c_v$ should have a value closer to $c_v \sim 0.1$ [15] [16] . If this consensus value of $c_v$ is applied to the STARM model, it will not be possible to find Maxwell-like linear rheology under any conditions. However, this is in conflict with real experimental observations of Maxwell-like linear rheology in wormlike micelles [20]. Although the STARM model is derived for the fast breaking CLF regime, this loss of Maxwell-like rheology comes from constraint release and will also emerge when one considers the fast breaking reptation limit. It is not immediately clear whether this conflict primarily points to new physics (e.g. different $c_v$ for flexible vs semi-flexible polymers) or problems with the STARM model itself (e.g. the continuum approximation of Rouse motions). Additional discussion is provided in the following subsections.

## 3.2. A closer look at constraint release: new physics

The standard value of $c_v \sim 0.1$ emerged via studies of constitutive modelling for flexible, unbreaking, and monodisperse linear chain polymers [15] [16]. For any polymer system meeting these descriptors (independent of chemistry) it is believed that the same value of $c_v$ should apply. Wormlike micelles are linear chain polymers, but they are also semi-flexible living polymers with a polydisperse molecular weight distribution – therefore, it is not immediately obvious that the same 'universal' value of $c_v$ should apply.

For example, one can argue that larger values of $c_v > 0.1$ might be possible for some systems of living polymers. In unbreaking polymers, there is an a-priori upper bound of $c_v < 1$ based on the fact that chains cannot freely pass through their entanglements [16]. However, a living polymer could effectively pass through an entanglement by breaking apart and recombining on the other side. Larger values of $c_\nu$ may be therefore be possible in living polymer systems, but the physics described here seem unlikely (or at least unimportant) for wormlike micelles in light of the fact that we are interested in explaining *smaller* apparent values of $c_\nu$.

To explain a smaller value of $c_\nu$ in wormlike micelles, we consider the effects of semi-flexibility – in particular, we consider the effect of $\alpha_e$ (the number of persistence lengths per tube segment). When $\alpha_e \gg 1$, entanglement segments are much longer than the persistence length and the intrinsic stiffness of the micelles should not have a significant effect. However, for $\alpha_e \ll 1$ the persistence length will span several tube segments and this may lead to substantially different physics for constraint release phenomena.

The STARM model partially accounts for the effect of varying $\alpha_e$ in terms of a modification to the equilibrium tangent correlation tensor. Rescaling the STARM model to be written in terms of $\bar{p} = \alpha_e p$, we see that (ignoring terms from equation (79)) varying $\alpha_e$ results in a rescaling of the constraint release term, $c_\nu^{eff} = \alpha_e^2 c_\nu$. Thus, the STARM model already partially accounts for a reduced effectiveness of constraint release attributed to semi-flexibility whenever $\alpha_e < 1$. However, to achieve a Maxwell-like linear rheology with $c_\nu^{eff} \sim 10^{-3}$ while maintaining a 'true' value of $c_\nu = 0.1$, one would need $\alpha_e = 0.1$, implying that a single persistence length spans ten tube segments! Given the typical diameter (4nm) and persistence length ($20 - 50$ nm) of a micelle, it is geometrically impossible to fit ten tube segments along a single persistence length, especially when the micelle volume fraction is below ten percent. Therefore, to obtain small values of $c_\nu^{eff}$ with reasonable values of $\alpha_e$ one must consider ways in which semiflexibility impact the 'true' value of $c_\nu$.

To that end, we would like to present a simplified picture of constraint release for estimating the 'true' $c_\nu$ in flexible and semi-flexible polymers. This analysis is not intended to produce a precise formula for $c_\nu$ but rather to provide tools for thinking more broadly about underlying factors at play.

Supposing that a single tube segment (length equal to diameter) experiences topological interactions with $N$ neighboring chains, but restrictions on its lateral movements are primarily attributed to a subset $N_C$ of those chains, on average. If one of those $N_C$ chains is released, the entanglement segment will still be held in place by the remaining $N_C - 1$ chains, provided $N_C > 1$.

If each polymer takes an average time $\tau$ to vacate a tube segment, then the typical time to add or remove topological interactions with one of the $N$ neighboring chains is $\tau/N$. Since the release of all constraints is a very rare event, with each update the probability that a tube segment has finally lost all its active constraints is roughly $(1 - N_C/N)^N$ or approximately $\exp(-N_C)$ whenever $N \gg N_C$. Thus, the constraint release frequency goes as $\nu = N \exp(-N_C)/\tau$, which implies $c_\nu = \tau \nu = N \exp(-N_C)$. For flexible polymers, where $N \sim 20$ [27] and $c_\nu \sim 0.1$, our simplified picture of constraint release implies that the effective tube surrounding an entanglement segment is defined by topological interactions from about $N_C \sim 5$ neighboring chains.

In the limit of highly entangled monodisperse rod-like polymers, the stiffness of the rod itself should reduce the effectiveness of constraint release – if the entanglements along a portion of the chain are all released simultaneously, that portion of the rod is not able to update its orientation because it is still held in place by entanglements elsewhere along the rod (c.f. Figure *12*). In this limit, the simplified model suggests that $c_\nu \sim N/\alpha_e \exp(-N_C/\alpha_e)$ where $\alpha_e$ is the number of tube segments per persistence length (or per rod, in this case). Thus, for $\alpha_e < 1$, constraint release may be less important than constraint release in well entangled flexible polymers ($c_\nu = 0.1$). In particular, using $N = 20$ and $N_C = 5$, we find that $c_\nu^{eff} = \alpha_e^2 c_\nu = 10^{-3}$ for $\alpha_e \sim 0.5$, which is not geometrically impossible for wormlike micelles. If this hypothesis is correct, one would also expect to see effects of an anomalously small $c_\nu$ in the linear and non-linear rheology of unbreakable but semi-stiff entangled linear polymers. However, it should be noted that well entangled rigid rods can rotate via "cooperative interactions" with neighboring rods [28] – including these effects will lead to a more complex picture of constraints in both flexible and semiflexible entangled polymers, and is unfortunately beyond the scope of the present report.

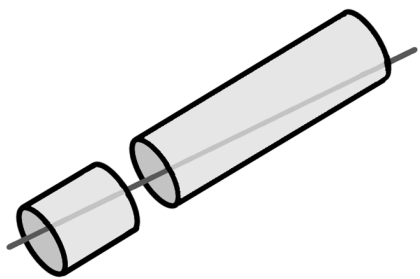

*Figure 12: Cartoon illustrating the weak effectiveness of constraint release relaxation in rigid rods – eliminating the constraints about an isolated portion of the tube does nothing to increase the range of rotational motion available to the rod.*

To review, we have suggested two different mechanisms by which semi-flexibility might reduce the effectiveness of constraint release: (1) by increasing the decay length embedded in $\boldsymbol{f}^{eq}$ and (2) by keeping tube segments 'trapped' in their current orientation even when all local constraints have been released. It is our view that the latter should be dominant whenever the persistence length becomes comparable to the tube diameter.

We would like to point out that our argument for a decreased value of $c_\nu$ is distinct from the arguments of Milner et. al., who suggested that constraint release could be marginally less effective for living polymers (e.g. due to 'shuffling' effects) even if the same value of $c_\nu$ were to apply [16]. However, from our calculations we find that shear banding is found in the STARM model only when $c_\nu < 0.1$, which agrees with the reported values from studies on unbreaking monodisperse polymers [15] [16]. Therefore, it seems that shuffling alone probably doesn't lead to a significant change in the effectiveness of constraint release.

### 3.3. A closer look at constraint release: conflicting approximations?

Although there are physical grounds to suggest that the effect of constraint release will be weaker for semi-flexible polymers, it may be that this explanation is not suitable (or not sufficient) to explain Maxwell like linear rheology in all living polymer systems where it has been observed [29]. To that end, we also consider potential weaknesses in the STARM model.

One improvement of the STARM model over the previously published implementation of the MML model [16] is that the STARM model does not rely on an arbitrary truncation of a Fourier series to ensure a finite size of the tube diameter. However, one previously overlooked feature of the MML approach that in truncating the Fourier series representation of the equilibrium tangent correlation tensor, it simultaneously truncates the Rouse relaxation spectrum of the tube.

In the STARM model, the typical constraint release time is equivalent to the terminal relaxation time by breaking/CLF – therefore, one might argue that the effects from constraint release Rouse motion should not be dominant on timescales faster than the terminal relaxation time. This would hold true for the MML model approach, where the tube has a fastest Rouse relaxation time because of the Fourier series truncation. However, it is not true in the STARM model – constraint release dominates the high frequency response (c.f. Figure 10).

In numerical calculations of the STARM model, a cutoff Rouse mode is always present based on how the problem has been discretized in $p$ (c.f. Figure 30). Usually, this cutoff Rouse mode is pushed out of range by mesh refinement, but there may be physical grounds to argue that more realistic predictions are obtained when the mesh is not converged but instead carefully discretized to preserve a particular spectrum of constraint release relaxation times. If this is the correct explanation, then many of the leading full-chain tube based models of non-linear polymer rheology will need to revisit the way in which constraint release phenomena have been modelled [15] [25].

In our view, constitutive modelling of constraint release phenomena has not yet been studied at a sufficient level of detail to resolve these questions precisely. It may be that a better understanding can be obtained via a study of stress relaxation in slip link models with reversible scission reactions. In any case, there may be future refinements to the way in which constraint release is implemented, but for now we will proceed to consider the STARM model as written. The calculations of nonlinear rheology in the following sections are fully converged in both $t$ and $p$.

## 3.2. Nonlinear rheology of the STARM model

### 3.2.1. Steady Simple Shear Flow

Turning now to the predictions for the nonlinear rheology of wormlike micelles, we focus our study on the effects of varying $\theta, c_\nu$, and the Weissenberg number $Wi$. Note that the Weissenberg number, $Wi$, is just the shear rate $\dot\gamma$ or strain rate $\dot\epsilon$ scaled by a characteristic stress relaxation time. We do not vary any of the dimensional parameters, $G_e, \tau$, or $\tau_S$: $G_e$ rescales the stress, $\tau$ rescales time, and $\tau_S$ can be inferred from $\tau$ and $\theta$, as discussed in section 2.7. We also fix $\alpha_e = 1$, since varying $\alpha_e$ has the same effect as varying $c_\nu$. Finally, we remind the reader that all calculations pertain to living polymers in the fast breaking CLF regime ($\zeta \ll 1$, $\zeta_R \ll 1$, and $\zeta_R \bar{Z}^2 > 1$).

We begin with steady simple shear flow. In Figure 13, we present steady state shear stresses over a range of Weissenberg numbers for $c_\nu = 0.1$ and $\theta = 10^0, 10^1, 10^2$, and $10^3$. The choice of characteristic stress relaxation time for the Weissenberg number is made to be the true terminal relaxation time $\tau_0 = \lim_{\omega \to 0} G'(\omega)/\omega G''(\omega)$ – this ensures that the predictions for shear stresses and normal stresses (when scaled by the compliance modulus, $J_e^0 = \lim_{\omega \to 0} G''(\omega)/G'(\omega)^2$) all collapse for $Wi \ll 1$. Showing the model predictions in this way makes it easier to unambiguously identify changes in the nonlinear rheology predictions.

Whereas $\theta$ has very little impact on the linear rheology, Figure 13 shows that $\theta$ is of considerable importance to the nonlinear rheology because it sets the timescale for stretch relaxation and the amount of shear thinning that one sees for shear rates above $Wi > 1$.

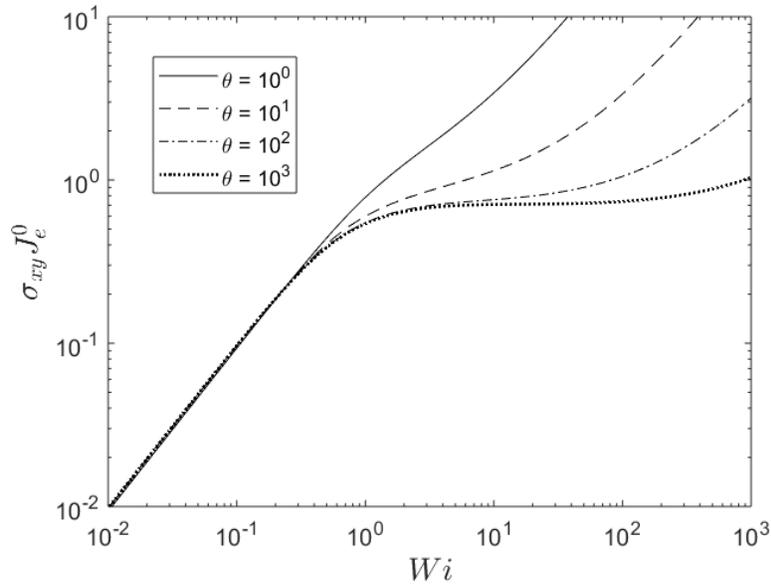

*Figure 13:  Predictions for the steady shear stress, $\sigma_{xy}$, of the STARM model with $c_\nu = 0.1$ and $\theta$ varying from 1 to 1000.  We observe increased shear thinning with increasing values of $\theta$.*

The predictions of Figure 13 are very similar to what has been previously reported theoretically for the rheology of well-entangled unbreaking linear polymers in steady simple shear flow:  increasing $\theta = \tau_{rep}/\tau_R$ leads to increased shear thinning in the range of $1 < Wi < \theta$, and if constraint release is sufficiently strong, $c_\nu \sim 0.1$, no shear banding is predicted for $\theta < 1000$ [15] [16] (i.e. the stress is always increasing with increasing shear rate).  Even in the nonlinear rheology, the main effects of reversible scission appear to be (1) a rescaling of the terminal relaxation time $\tau$ and (2) a narrowing of the relaxation spectra for reptation and chain retraction.

Compared to other models of well-entangled polymers, Figure 13 reveals a weakness of the STARM model at shear rates $Wi \gg \theta$, where (in contrast to experimental observations) there is no shear thinning.  This could be amended, as it was in the GLaMM model, by suppressing CCR whenever chains are stretched.  For the time being, however, it is our view that such a fix is highly speculative – we prefer to rely on well-established physics and admit that the STARM model as presently formulated is not suitable when chains are highly stretched by flow, $Wi \gg \theta$.  Further improvements to the constitutive modelling of well-entangled and unbreaking polymers are needed to inform improvements to the STARM model under such conditions.

Next, Figure 14 considers the effect of varying $c_\nu$ for fixed $\theta = 100$.  As discussed in section 3.23.2, one possible interpretation for the varied values of $c_\nu$ is that smaller values of $c_\nu < 10^{-2}$ may be applicable to semi-flexible chains (such as wormlike micelles) and larger values $c_\nu \sim 0.1$ are well-suited to unbreaking flexible polymers.  As established in the literature on monodisperse unbreaking polymers [15] [16], we see that smaller values of $c_\nu$ can lead to shear banding instabilities (i.e. a negative slope of the flow curve) for sufficiently large values of $\theta$.  The mechanism for shear banding shown here is that of Reptation Reaction model, where tube segments over-rotate into the direction of flow [17] [18] : the STARM model is not suitable for modelling shear banding as a flow-induced nematic phase transition [5]. However, we remind the reader that whereas the Reptation Reaction model was developed for the fast breaking reptation regime, the depiction of non-linear relaxation processes in the STARM model are best suited to the fast breaking CLF regime.

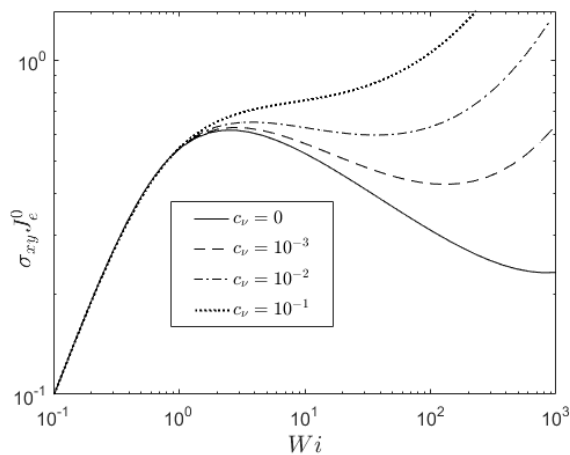

Figure 14: Predictions for the steady shear stress, $\sigma_{xy}$, of the STARM model with varying $c_v$ and $\theta = 100$. Note that shear banding is possible for values of $c_v$ below $c_v = 0.1$, which is the value typically assigned to well entangled flexible linear polymers. The shear stress and shear rates have been scaled with the compliance modulus, $J_e^0$, and terminal relaxation time, respectively, so that all curves collapse for $Wi \ll 1$.

While Figure 14 makes it clear that small values of $c_v$ are needed for shear banding, shear banding also requires a modest value of $\theta$ rather than a small one. With smaller values of $\theta$, chains become stretched (increasing their stress) even as they rotate towards the direction of flow. In Figure 15, we present predictions for the shear stress with $c_v = 0$ and varying $\theta$. Comparing Figure 15 and Figure 11, we see that all shear banding instances of the STARM model have a nearly Maxwellian linear rheology, but the converse is not true – for $\theta < 10$, the flow curve is always increasing and the linear rheology is Maxwell-like. Experimentally constructed 'flow phase diagrams' for systems with Maxwell-like linear rheology (i.e. $c_v = 0$) have shown a similar transition into shear banding with increasing surfactant concentration or decreasing temperature [5]. In those experiments, however, the nonlinear aspects of the shear banding instability were connected to a flow-induced phase transition to nematic order, which the STARM model is not equipped to describe.

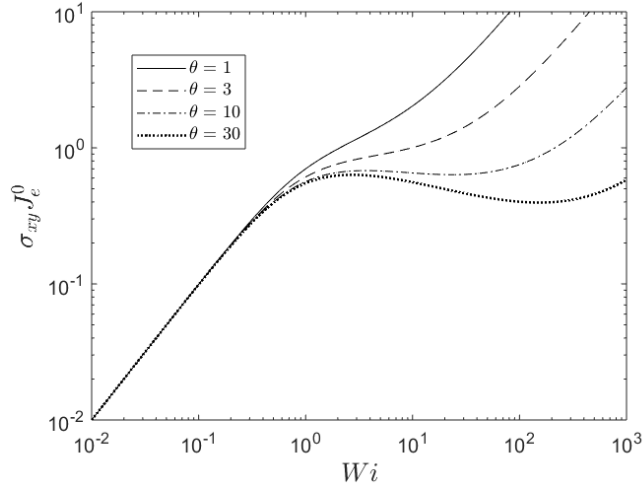

*Figure 15: Predictions for the steady shear stress, $\sigma_{xy}$, of the STARM model with $c_\nu = 0$ and $\theta$ varying from 1 to 1000. Negative sloping curves (leading to shear banding) are only observed for sufficiently large $\theta$, beginning around $\theta \sim 10$.*

Finally, we consider how varying $\theta$ and $c_\nu$ affects predictions the first normal stress difference. In Figure 16 we fix $c_\nu = 0.1$ and vary $\theta = 10^0, 10^1, 10^2, 10^3$, while in Figure 17 we fix $\theta = 100$ and vary $c_\nu = 0, 10^{-3}, 10^{-2}, 10^{-1}$. The predictions are similar to what is known for well entangled unbreaking polymer systems: increasing $\theta$ and decreasing $c_\nu$ lead to 'shear thinning' of the normal stresses in the range of $1 < Wi < \theta$. The effect of varying $c_\nu$ appears to be more important for shear stresses than normal stresses. In particular, Figure 14 shows that decreasing $c_\nu$ controls a transition from stable homogeneous flow (monotonic flow curve) to shear banded flow (non-monotonic flow curve) and there is no counterpart observed in Figure 17. Recall also that when shear banding instabilities occur, unstable portions of the homogeneous flow curve are not accessible experimentally – the shear stresses exhibit a plateau and the normal stress differences exhibit a linear ramp [17].

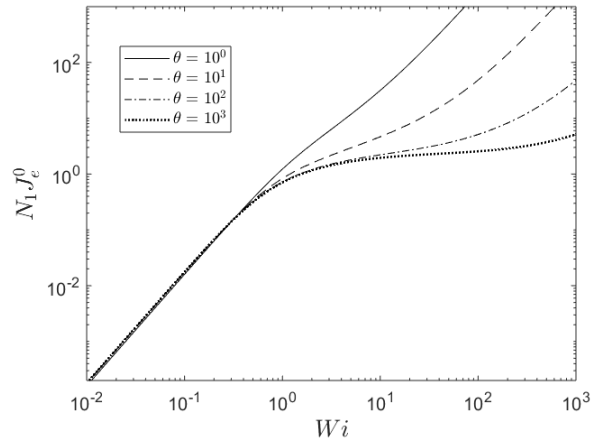

*Figure 16: Predictions for the first normal stress difference, $N_1$, in steady simple shear flow for the STARM model with $c_\nu = 0.1$ and $\theta$ varying from 1 to 1000. We observe increased shear thinning with increasing values of $\theta$, and none of these flow curves are subject to shear banding (c.f. Figure 13).*

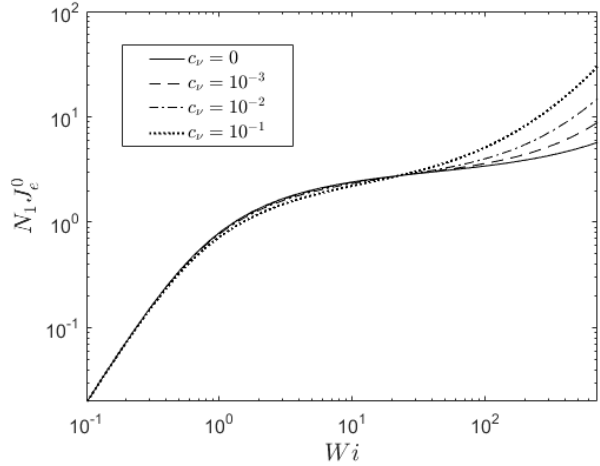

*Figure 17: Predictions for the steady shear first normal stress difference, $N_1$, of the STARM model with varying $c_\nu$ and $\theta = 100$. The normal stress and shear rates have been scaled with the compliance modulus, $J_e^0$, and terminal relaxation time, respectively, so that all curves collapse for $Wi \ll 1$. For this family of flow curves, shear banding instabilities will develop for $c_\nu \leq 10^{-2}$.*

### 3.2.2. Steady uniaxial extension

Next, we consider predictions for the STARM model under steady uniaxial extension with strain rate $\dot{\epsilon}$ and $Wi = \dot{\epsilon}\tau_0$. First, we fix $c_\nu = 0.1$ and vary $\theta = 10^0, 10^1, 10^2, 10^3$. For $Wi \sim \theta$, chains are stretched faster than they can retract within the tube and since we have not yet accounted for finite extensibility the extensional viscosity diverges. To keep this non-physical divergence out of view, we limit our reported results to flows in which the steady state stretch $\bar{\lambda}$ remains below an arbitrary cutoff of $\bar{\lambda} < 2$. Figure 18 shows the first normal stress, $N_1$, as a function of the Weissenberg number $Wi$ of the deformation for varying $\theta$ and fixed $c_\nu = 0.1$. As in Figure 13, we see that increasing $\theta$ leads to increasing shear softening (or reduced strain hardening) in the range of $1 < Wi < \theta$.

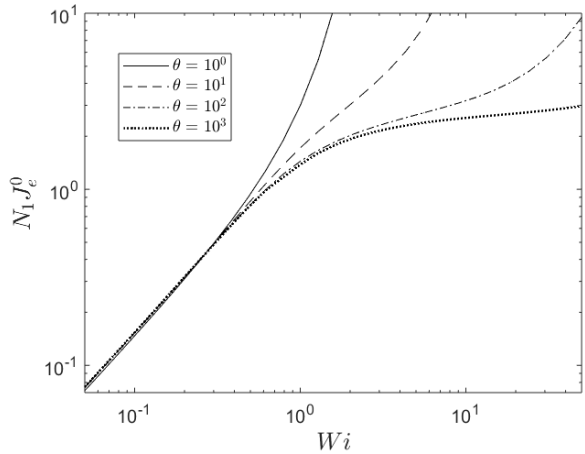

*Figure 18: Predictions for the first normal stress difference, $N_1$, in uniaxial extensional flow for the STARM model with $c_\nu = 0.1$ and $\theta$ varying from 1 to 1000. Note that we observe increased shear thinning with increasing values of $\theta$.*

Figure 19 considers the same uniaxial extensional flow with fixed $\theta = 100$ and varying $c_\nu = 0, 10^{-3}, 10^{-2}, 10^{-1}$. Once again, we see that decreasing $c_\nu$ leads to increasing shear thinning, however the effect is much weaker than it was in simple shear flow. Even at $c_\nu = 0$ there is no opportunity for a 'necking' instability (the extensional flow analogue to shear banding).

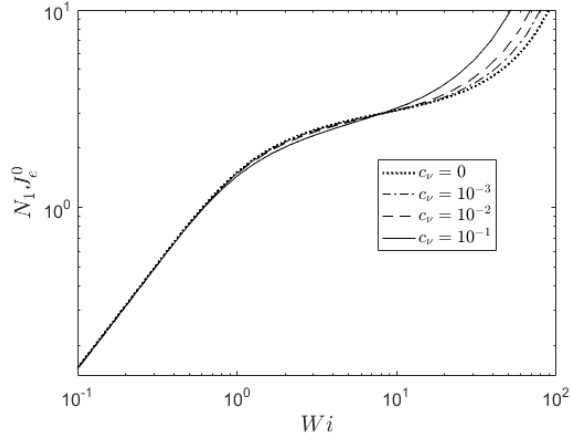

*Figure 19: Predictions for the first normal stress difference, $N_1$, in steady uniaxial extensional flow for the STARM model with fixed $\theta = 100$ and varying $c_\nu$. We observe a slight increase in shear thinning with decreasing $c_\nu$.*

### 3.2.3. Startup transients: shear and extension

Having described the predictions of the STARM model under steady flow conditions, we will briefly describe the transient rheological response during start-up of steady shear and extension in a system that begins from equilibrium. We limit our survey of start-up transients systems a single value of $\theta = 100$ and two values of $c_\nu = 0, 0.1$. In Figure 20(a), we see a small stress overshoot for shear rates exceeding $Wi \sim 1$, and the stress overshoot is larger in Figure 20(b) partly due to the increased shear thinning that occurs at smaller values of $c_\nu$. However, for Figure 20(b) the system is likely to shear band in steady state at the two highest $Wi$, which will lead to changes in the transient rheological signature.

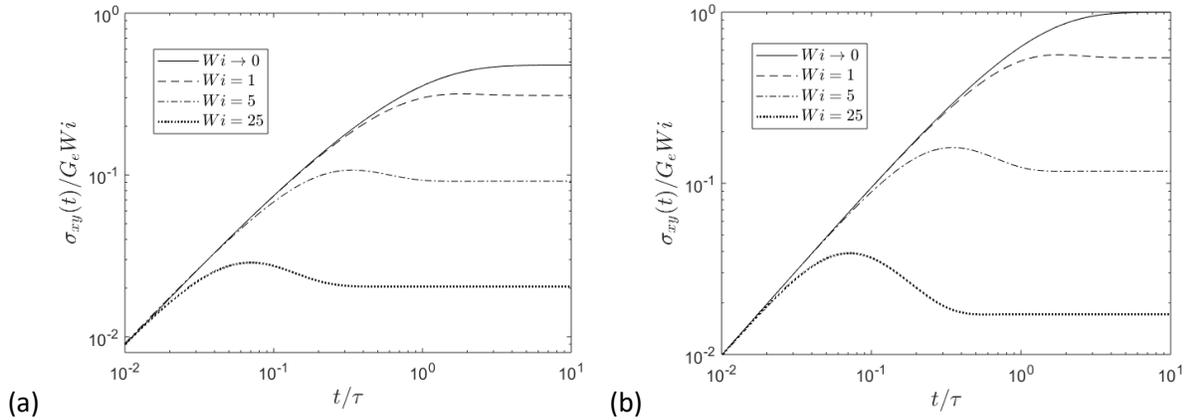

(a)                                                                    (b)

*Figure 20: Predictions for the shear stress, $\sigma_{xy}$, in start-up shear flow for the STARM model with $\theta = 100$ and (a) $c_\nu = 0.1$, (b) $c_\nu = 0$. We observe a small stress overshoot for shear rates exceeding $Wi \sim 1$. The stress overshoot is larger for lower values of $c_\nu$.*

In Figure 21, we consider start-up transients for extensional flows, using $\theta = 100$ and $c_\nu = 0.1$ (results are similar at other $c_\nu$, c.f. Figure 19). For strain rates between $1/\tau$ and $1/\tau_S$ we see a decreasing steady state viscosity – this is because chains have oriented into the direction of extension but are not yet stretching. For strain rates exceeding $1/\tau_S$ we see runaway strain hardening. Once again, all of the results reported in Figure 21 consider only extension rates and/or times for which the stretch is not very large, $\bar{\lambda} < 2$.

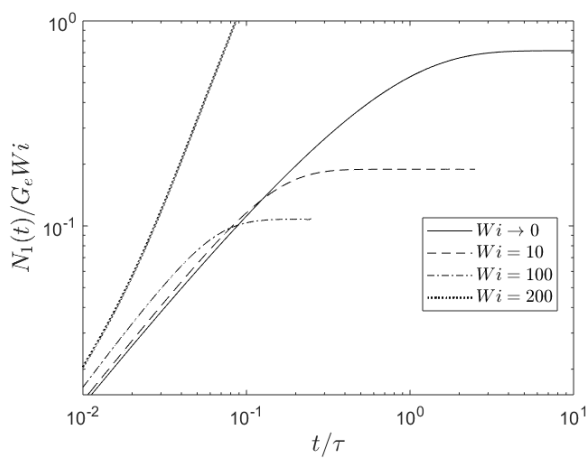

*Figure 21: Predictions for the first normal stress difference, $N_1$, in start-up uniaxial extensional flow for the STARM model with fixed $\theta = 100$ and $c_v = 0.1$. We observe the onset of strain hardening at strain rates exceeding $Wi \sim \theta$.*

# 4. Comparison to experimental systems

Finally, we now pursue a direct quantitative comparison to experimental data. In particular, we will compare to rheological data (linear rheology, shear stresses, and first normal stress differences) reported in the recent work by Hwang et. al [30]. This study considered wormlike micelles formed by CPyCl (cetylpyridinium chloride) at concentrations of 95mM with salt-to-surfactant ratios $R = [\text{NaCl}]/[\text{CPyCl}] = 0.55$ and $0.79$. Both of these materials appear to be well entangled, and in our view neither is convincingly shear banding in its nonlinear rheology[6].

The sample at lower salinity may not be in the fast breaking CLF limit, and we will discuss how this becomes apparent during the fitting process. Based on its linear rheology, the sample at higher salinity does appear to be fast breaking, but the authors suggest it may also have a branched architecture. However, the branched structure is likely quite tenuous [31] and as such may not have a dominant effect on the system's rheology. The rheological data is not sufficient to establish whether the system is fast breaking by reptation or CLF, but here we will assume that a fast breaking CLF model can be applied.

Our choice of experimental systems specifically attempts to avoid materials that shear band, and this is necessary for the time being: the STARM model does not account for the effect of a velocity gradient that changes over length-scales comparable to the size of a single chain, and as such the model predictions in steady simple shear flow for shear banding materials remain ill-posed [14]. The development of leading order nonlocal or gradient corrections to the constitutive model will be a topic for future work.

---

[6] Hwang et. al. define the R = 0.79 sample as 'shear banding' for a range of shear rates where the shear stress is observed to be *relatively* constant. However, this classification is subjective and in context with prior work by Berret et. al. [5], we would tentatively classify the R = 0.79 sample as being very close to but distinctly outside of the shear banding regime. To support this classification, we point out that the stress 'plateau' of the R = 0.79 sample covers less than a decade of shear rates, and the transition into and out of the stress plateau is smooth, with no clear evidence of an underlying discontinuity in the value/slope of the flow curve.

The results of this section are summarized as follows: In section 4.1, we fit the STARM model to the linear rheology and non-linear shear rheology (including both shear and first stress differences). The STARM model is found to work well for the high salinity material, $R = 0.79$, but it performs poorly by comparison for the low salinity material, $R = 0.55$. Such a disparity would be expected if, for example, the high salinity material were in the fast breaking CLF regime and the low salinity sample were slow breaking. We pursue this hypothesis in two ways:

First, we present a phenomenological modification to the STARM model deigned to separate a potentially useful (albeit non-physical) description of thermal constraint release in slow breaking systems from a physical description of convective constraint release. This approximation is shown to improve fitting to shear rheology but has no advantage for fitting the first normal stress difference.

Second, we fit the linear and non-linear rheology of the low salinity samples using constitutive models that allow for slow breaking times. Specifically, we show that the linear rheology is described well by the 'toy' shuffling model given earlier in this report, and the shear rheology data is fit well by the Living Rolie Poly model [11]. Once again, however, a poor fit is obtained for the first normal stress data.

Finally, as a point of reference we fit the non-linear rheology of the high and low salinity samples using a Giesekus constitutive equation, which is a popular semi-empirical viscoelastic model for well entangled WLM. We show that, as with the STARM model, the Giesekus model performs well for fitting shear stresses but performs poorly for fitting normal stresses.

Overall, we find that the STARM model appears to work well when it is applied to a system that is well entangled and fast breaking, which is consistent with the assumptions of the model. When the STARM model is applied to a system that does not appear to be fast breaking, we find that a small phenomenological modification to equation $(81)$ allows the STARM model to match the predictive power of a model designed for use outside the fast breaking limit.

## 4.1 STARM Fits

In the comparisons considered here, we begin in Figure 22 by fitting STARM to the reported linear rheology data. At high salinity, $R = 0.79$, a Maxwell-like response is seen in the linear rheology, and the STARM model provides a good fit for any value of $c_v < 10^{-3}$ (as discussed in preceding sections, this value is much lower than the 'typical' value assigned to unbreaking flexible polymers). At lower salinity, $R = 0.55$, the linear rheology becomes non-Maxwell and larger values of $c_v \sim 1.7$ are required to obtain a good fit. However, there are two problems with this fitted value: First, it is not clear why the value of $c_v$ should have *any dependence* on the salinity – salinity has a large effect on the length of a micelle and a small effect on its persistence length, but there is no clear mechanism by which it might effect such a large change in the effectiveness of constraint release. Second, and perhaps more importantly, values of $c_v > 1$ are generally considered out-of-bounds when polymers are well entangled and cannot pass through one another[7] [15] [16]. For the surfactant chemistry considered here, fitting to the linear rheology of the high salt sample suggests that $c_v < 10^{-3}$ is the correct value for all salt ratios where the system is well entangled, and larger apparent values of $c_v$ identify systems that are not fast breaking.

---

[7] For living polymers that can reversibly break or branch, interpretations of $c_v > 1$ are possible but in our view not likely to be correct, at least for this particular case. Given that this system exhibits $c_v < 10^{-3}$ at high salt concentrations, it is likely that the low salinity sample has simply become slow breaking (e.g. because chains are shorter) and not undergone a radical change in its constraint release processes.

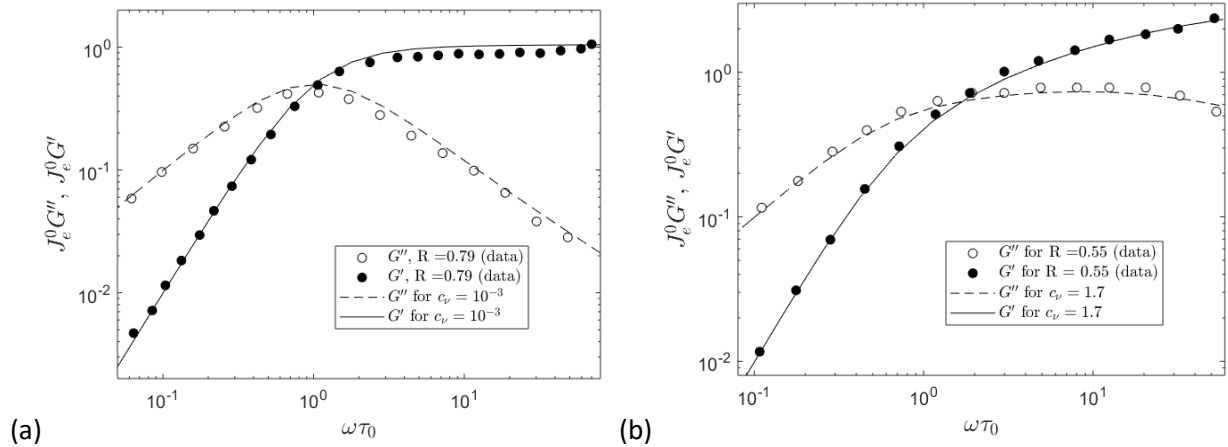

*Figure 22: Comparing experimental data for the loss modulus (empty circles) and storage modulus (filled circles) with fits obtained via STARM (curves). The measurements correspond to 95mM CPCl in a NaCl brine with salt-to-surfactant ratios of (a) R = 0.79 and (b) R = 0.55.*

Turning now to the nonlinear rheology data, Figure 23 shows that the STARM model fit to steady state shear stress looks much better for the high salt sample, $R = 0.79$ than the low salt sample, $R = 0.55$. In the high salt sample, the STARM model departs from the rheological data only for shear rates exceeding $Wi > 10$, but at such high $Wi$ it is possible that confounding factors in the experiments (e.g. viscoelastic instabilities) might be contributing to the observed disparity. In the low salt sample, there is a noticeable difference for the entire range of interest, $Wi > 1$. We have previously remarked that values of $c_v > 0.1$ are physically unrealistic, so it is not surprising to see that nonlinear rheology predictions for the low salt sample are likewise unrealistic. The linear rheology for the low salt sample shows a good fit, but that good fit probably leverages the wrong physics (constraint release rather than slow breaking times). Therefore, the poor fit shown in Figure 23b corroborates the hypothesis that large values of fitted $c_v$ identify samples are not well described by a 'fast breaking' model, $\zeta \ll 1$.

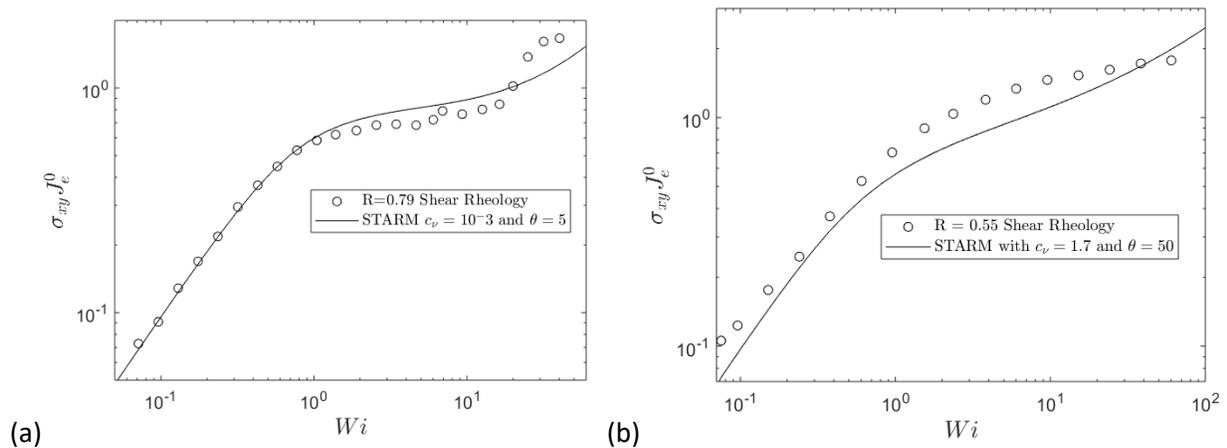

*Figure 23: Comparing experimental data for steady state shear stresses (circles) with fits obtained via STARM (curves). The measurements correspond to 95mM CPCl in a NaCl brine with salt-to-surfactant ratios of (a) R = 0.79 and (b) R = 0.55.*

Experimental measurements for the first normal stress difference are also reported by Hwang et. al., but there are some concerns that the linear and non-linear rheology data may not be self-consistent: the way our figures have been scaled, measured normal stresses should always converge to model predictions in the limit of $Wi \to 0$, but visual inspection of Figure 24 shows an apparent offset in both cases. If the data for complex moduli and shear stresses were taken in a Couette flow and the data for normal stresses were taken from a cone/plate flow, the difficulty in comparing those data sets could partially explain this observation (the authors do not report details on what flow geometries were used for each data set). In any case, the discussion will focus only on qualitative aspects of the comparison between experimental measurements and model predictions.

Considering the high salt sample, Figure 24(a) shows there is qualitative agreement between the experimental observations and the model predictions of a first normal stress difference – in particular, the model and the data would nicely superimpose if the experimentally measured stresses were multiplied by a factor of about three. In the low salt sample, qualitative agreement is quite poor by comparison – no vertical shift would lead the data to superimpose with the model.

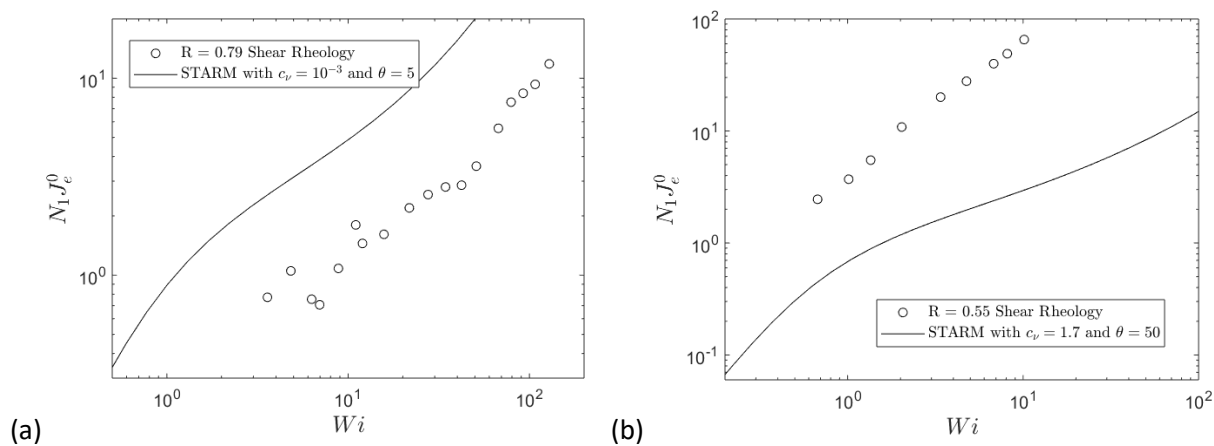

(a)                               (b)

*Figure 24: Comparing experimental data for steady first normal stresses (circles) with fits obtained via STARM (curves). The measurements correspond to 95mM CPCl in a NaCl brine with salt-to-surfactant ratios of (a) R = 0.79 and (b) R = 0.55.*

In our view, the low salinity sample, $R = 0.55$, is unlikely to be fast breaking (neither $\zeta_R \ll 1$ nor $\zeta \ll 1$) and will not be described well by any model built on a fast-breaking approximation. Instead, it is likely that a more realistic description can be obtained by allowing longer timescales for breaking or 'shuffling' of the chains – this will be discussed in the next subsection. However, given that the linear rheology fit is quite good, we briefly offer a phenomenological strategy for improving the fits to nonlinear rheology data.

When the STARM model is fit to the linear rheology of a sample that is not fast breaking, one obtains a value for $c_v$ that is unphysically large, since dispersion in the reptation spectra is described via an overly-active constraint release process[8]. However, when the STARM model is subsequently applied to model nonlinear flows with convective constraint release, the depiction of constraint release as overly-active is not physical. Therefore, a first step for adapting the STARM model to systems that are not fast breaking is to revise equation (81) allowing separate values of $c_\nu$ for linear and nonlinear relaxation processes with only the latter constrained to realistic values:

$$\nu = \left( \frac{c_\nu^{(1)}}{\tau_{CLF}} + \frac{c_\nu^{(2)}}{\tau_s} \left( 1 - \frac{1}{\bar{\bar{\lambda}}} \right) \right) \qquad (95)$$

---

[8] Using the physics of constraint release to approximate the physics of reptation is not without precedent in the constitutive modelling of entangled linear polymers. We would suggest that the double reptation approximation (where the physics of reptation is used to approximate the effects of constraint release) makes the same connection in the opposite direction (des Cloizeaux, 1988 ).

Using $c_\nu^{(1)} = 1.7$ and $c_\nu^{(2)} = 10^{-3}$, we find that when comparing Figure 25 with Figure 23 (b) and Figure 24 (b), the modified STARM model gives an improved fit to the nonlinear rheology data of the R = 0.55 material, but mostly just with respect to the shear stresses. Predictions of normal stresses are still quite poor by comparision.

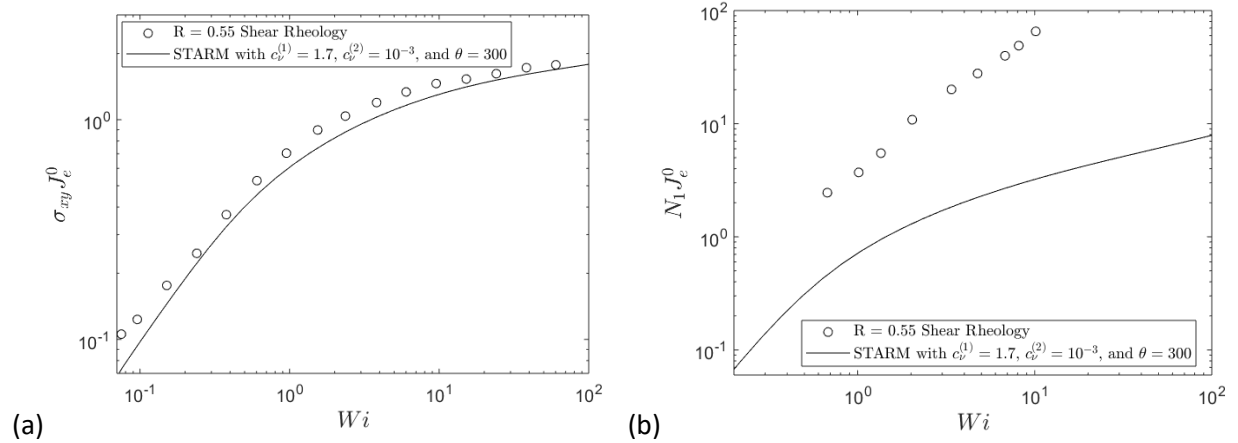

*Figure 25: Comparing experimental data (circles) for the steady state (a) shear stress and (b) first normal stress with STARM model predictions using different values of $c_\nu$ for the linear and nonlinear contributions to constraint release. The measurements correspond to 95mM CPCl in a NaCl brine with salt-to-surfactant ratio of R = 0.55.*

## 4.2 Fits by slow breaking models, $\zeta > 1$

To consider linear rheology prediction outside of any fast breaking limit, we will rely on the 'toy' shuffling approximation since its predictions are much much easier to compute and it is known to be valid in the extreme limits of $\zeta \ll 1$ and $\zeta \gg 1$.

We first show that the 'toy' shuffling model can fit the linear rheology data at $R = 0.55$ just as well as the STARM model does. Linear rheology predictions of the 'toy' shuffling model can be obtained for any value of $\zeta_{sh}$ and only require numerical evaluation of a single integral:

$$\frac{G^*(\omega)}{G_e} = i\omega \left[ \frac{1}{C(\omega, \zeta_{sh})} - \frac{1}{\zeta_{sh}} \right]^{-1} \qquad (96)$$

$$C \equiv \int_0^\infty dz \frac{z e^{-z}}{A/z} \left[ 1 - \frac{2}{z\sqrt{A}} \tanh\left( \frac{z}{2}\sqrt{A} \right) \right] \qquad (97)$$

$$A \equiv z\left( \zeta_{sh}^{-1} + i\omega \right) \qquad (98)$$

A derivation for the above relation is provided in the appendix. Incidentally, this result is equivalent to what was previously found by Granek and Cates (GC) [10], whose 'Poisson Renewal' model only seems to differ from our 'toy' shuffling model in just two respects: first, the GC model allowed a $z$-dependent shuffling time (which we have excluded on physical grounds) and second, the GC model was framed as an integral equation rather than a partial differential equation.

In Figure 26, we show that $\zeta_{sh} = 10$ gives a good fit for the linear relaxation spectra of the low salt sample.

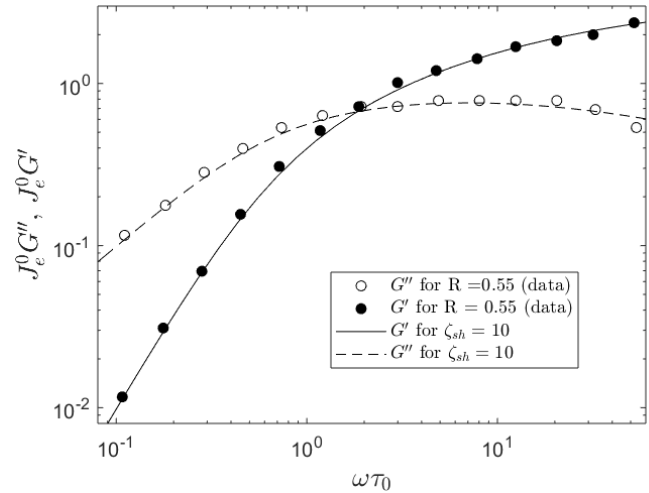

Figure 26: *Comparing experimental data for the loss modulus (empty circles) and storage modulus (filled circles) with a obtained from the 'toy' shuffling model of reptation and reversible scission. The measurements correspond to 95mM CPCl in a NaCl brine with salt-to-surfactant ratio of R = 0.55.*

Unfortunately, it is not possible to use the 'toy' shuffling model to directly make predictions for nonlinear rheology. For nonlinear rheology calculations outside of the fast breaking limit, one can instead make use of the 'Living Rolie Poly' (LRP) model to reasonably good effect [11].

An equally good fit to the linear rheology is possible in the LRP model (albeit with a significantly different estimate of $\zeta \sim 300$), and the subsequent nonlinear rheology predictions of the LRP model with $\zeta = 300$ shown in Figure 27 provide a more believable fit to experimental observations of steady state shear stresses when compared with the STARM predictions of Figure 23. For both shear stresses and normal stresses, the LRP fit is comparable to the results of Figure 25.

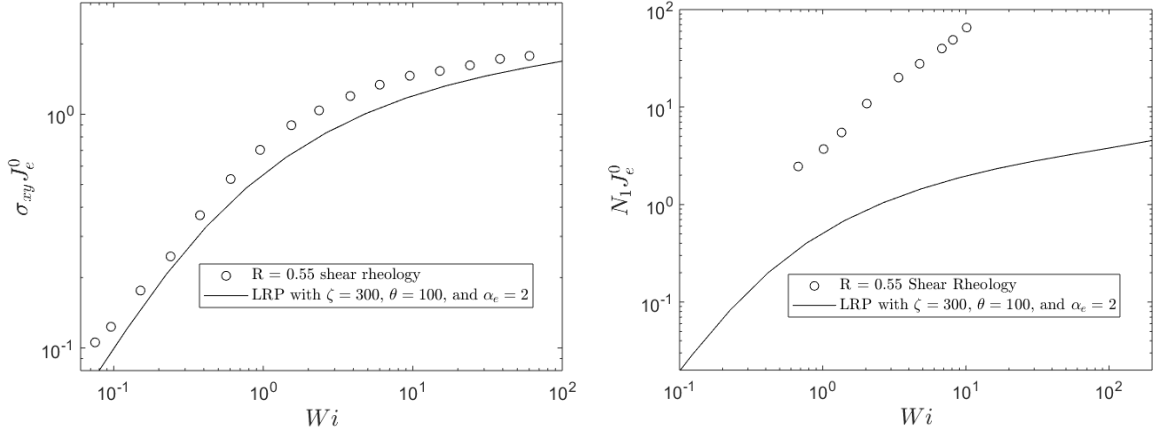

*Figure 27: Comparing experimental data for the (a) shear stresses and (b) first normal stresses with a fit from the 'Living Rolie Poly' model. The measurements correspond to 95mM CPCl in a NaCl brine with salt-to-surfactant ratio of R = 0.55.*

For the low salinity sample, descriptions of the first normal stress difference (whether by STARM or Living Rolie Poly) have been less accurate than descriptions of the linear rheology and shear stresses. It is our view that this is not necessarily a weakness of the population balance framework but could be an inherited weakness from the underlying constitutive models [15] [19], if only in the sense that the underlying constitutive models are better suited for flexible polymers than for semiflexible polymers. As and when improvements to the underlying constitutive models are developed, those improvements can be propagated through the appropriate population balance framework to obtain improved rheological predictions.

We do not pursue further here the issue of how to model the constitutive behavior of WLMs, allowing for CLF and CR, outside of the fast breaking CLF limit. However, we do plan to return to this topic in a future paper.

### 4.3 Fits by Giesekus model

As a point of reference, we also provide fits to the experimental data based on a Giesekus constitutive equation [32], for which the polymer contribution to the stress evolves by:

$$\overset{\triangledown}{\boldsymbol{\sigma}} = -\frac{1}{\tau}\boldsymbol{\sigma} - \left[\frac{\alpha}{G_e\tau}\right]\boldsymbol{\sigma}\cdot\boldsymbol{\sigma} \qquad (99)$$

The Giesekus model is a phenomenological constitutive equation for viscoelastic fluids, and it is a popular choice for fitting non-linear experimental data on wormlike micelles [6].

For the high salinity sample, $R = 0.79$, the Giesekus model gives a fit that is slightly better for simple shear flow and (in our opinion) considerably worse for the normal stresses:

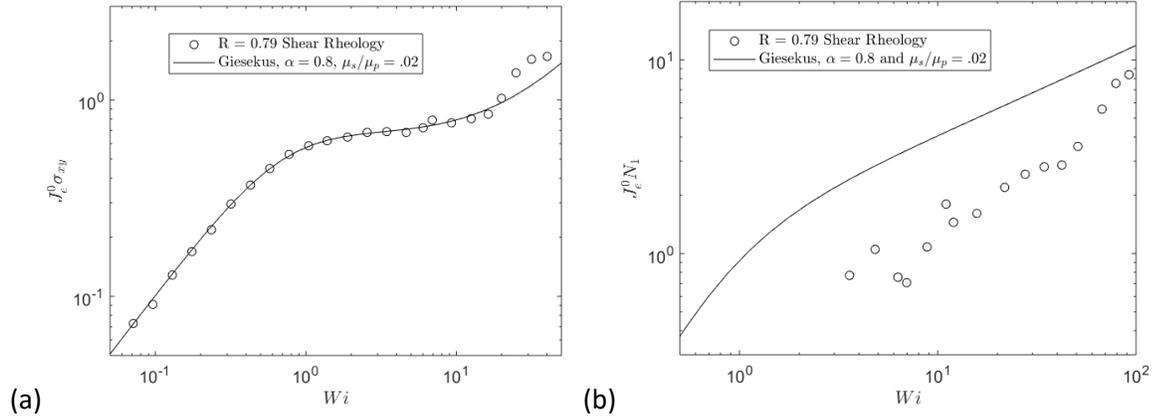

*Figure 28: Comparing experimental data for steady state shear stresses (a) and normal stresses (b) with fits obtained via the Giesekus model with $\alpha = 0.8$ and a solvent viscosity $\mu_S = 0.02 G\tau$. The measurements correspond to 95mM CPCl in a NaCl brine with a salt-to-surfactant ratios of R = 0.79.*

For the low salinity sample, $R = 0.55$, the Giesekus model (being a single-mode model) will provide a poor approximation of the linear rheology data. For the non-linear rheology data, however, the Giesekus model performs slightly better than the STARM model for both shear stresses and normal stresses. However, the comparison for normal stresses still leaves considerable room for improvement in our opinion.

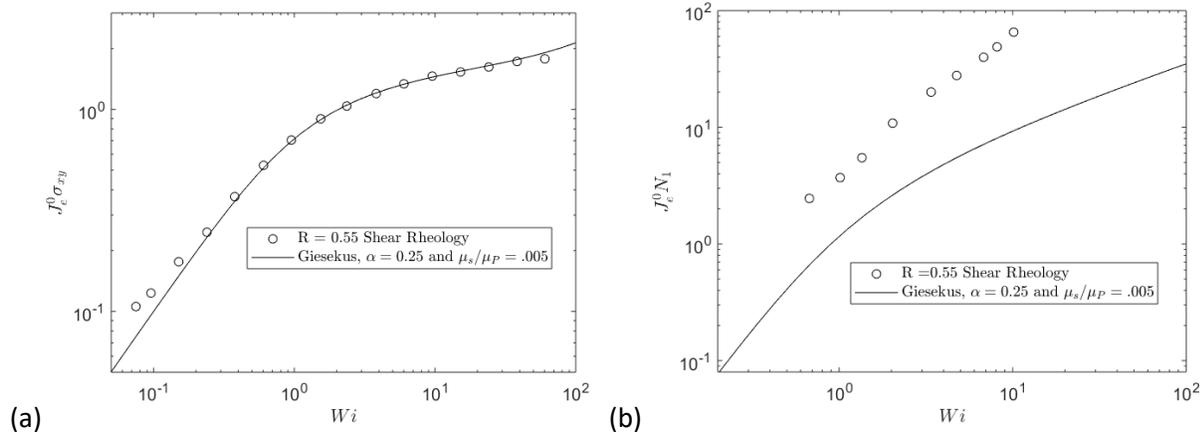

*Figure 29: Comparing experimental data for steady state shear stresses (a) and normal stresses (b) with fits obtained via the Giesekus model with $\alpha = 0.8$ and a solvent viscosity $\mu_S = 0.02 G\tau$. The measurements correspond to 95mM CPCl in a NaCl brine with a salt-to-surfactant ratios of R = 0.79.*

Overall, we find that the STARM model can provide predictions comparable to those of the Giesekus model for fast breaking systems. For slow-breaking systems, STARM can still perform well provided on permits a phenomenological re-interpretation of the thermal constraint release coefficient (c.f. equation (95) and Figure 25).

As the constitutive modelling tools for entangled polymers improve, the fitting power of the STARM model should advance beyond what the Giesekus model is able to attain. Absent that, however, the STARM model still has notable advantages; the Giesekus model is useful for fitting data but, unlike

STARM, the fit that one obtains does not provide any information about the material at the level of its microstructure.

## 5. Discussion and Conclusions

We have presented a new mathematical framework to describe stress relaxation dynamics in linear chain living polymers. In particular, we show that population balance equations can perform a detailed accounting for how the stretch, orientation, and alignment of entanglement segments are advected through the molecular weight distribution via the living polymerization reactions. When these population balances are combined with a suitable description for stress relaxation phenomena, one can systematically develop comprehensive linear and nonlinear constitutive equations for living polymers. We have used this approach to (1) confirm previously established results, (2) correct previously reported results, and (3) develop a more complete picture of nonlinear stress relaxation dynamics, incorporating chain retraction and convective constraint release.

Regarding (1), we confirm that in the fast breaking limits $\zeta \ll 1$ where stress relaxation is dominated by reptation ($\zeta_R \gg 1$) or contour length fluctuations ($\zeta_R \ll 1$ and $\zeta_R \bar{Z}^2 \gg 1$) stress relaxation is dominated by a single timescale $\tau$ that scales as $\tau \sim \zeta^{1/2}$ for the former and $\tau \sim \zeta^{1/2}\zeta_R^{1/4}$ for the latter [1]. Recall that $\bar{Z}$ is the mean entanglement number, $\zeta$ is the ratio of a typical breaking time to a typical reptation time, and $\zeta_R = 3\bar{Z}\zeta$ is the ratio of a typical breaking time to a typical chain's longest Rouse relaxation time.

Regarding (2), when chains are fast breaking and unentangled $\zeta_R \ll 1$ (or breaking is so fast that entanglements no longer matter, $\zeta_R \bar{Z}^2 \ll 1$) it was previously suggested that stress relaxation would occur primarily at chain ends with a single relaxation time scaling as $\tau \sim \zeta_R^{1/2}$ [1]. However, we have shown that in this limit most stress relaxation actually occurs in the chain interior, a result that was less well known but previously reported for liquid Selenium [9]. In our analysis, we have confirmed that this pathway for stress relaxation leads to a Rouse-like spectrum of stress relaxation times, the longest of which scales as $\tau \sim \zeta_R^{2/3}$.

Regarding (3), we provide the first derivation of a non-linear constitutive model including both chain retraction and CCR. Similar models were previously known to the literature (albeit with no substantive derivation) [16], but ours is the first to allow a finite stretch relaxation time and a natural cut-off length for the equilibrium tangent correlation tensor.

One significant advantage of the population balance modelling framework introduced in this report is that its utility is not restricted to constitutive modelling in the linear regime. To demonstrate that point, we have developed the STARM model for well entangled living polymers in the fast breaking CLF limit, $1/\bar{Z}^2 \ll \zeta_R \ll 1$. Evaluating the predictions of the STARM model and comparing to experimental observations of wormlike micelles, we find that the value of the constraint release parameter $c_\nu$ typically assigned to flexible unbreaking polymers $c_\nu \sim 0.1$ cannot be reconciled with the data for some systems of wormlike micelles, in which a single mode Maxwell is seen in the linear response. Instead, much smaller values, $c_\nu < 10^{-3}$, seem to be more appropriate. We have offered two interpretations for this anomalous result: (1) it may be that constraint release is a less effective stress relaxation process in semiflexible polymers or (2) the continuum description of Rouse-like motion common to modern tube-based constitutive models is not appropriate on very short timescales.

Whatever the underlying physical mechanism that leads to small apparent values of $c_v$ in wormlike micelles, it seems likely that a reduced effectiveness of constraint release in the linear relaxation spectra likely implies a reduced effectiveness of convective constraint release (CCR) in strong flow as well. Since CCR is thought to suppress shear banding in flexible unbreaking polymers [15] [16], it may be that the relative absence of CCR in wormlike micelles is responsible for the widespread experimental observations of shear banding phenomena in these systems. This line of reasoning can explain why the Reptation Reaction model [17] [18] (which ignores CCR entirely) is nonetheless a powerful predictor of shear banding instabilities in micellar systems.

Whereas the present report has largely focused on stress relaxation dynamics of living polymers in 'fast breaking' limits (e.g. $\zeta \ll 1$ and/or $\zeta_R \ll 1$), we have also briefly considered some preliminary strategies for linear and nonlinear constitutive modelling of wormlike micelles that are not fast breaking. Since many systems of practical interest are not fast breaking, we plan to return to this important topic in future work.

## 5. Acknowledgements

We would like to thank Daniel Read for valuable discussions in the early stages of this work. We would also like to thank Unilever and the CAFE4DM consortium for funding and technical input. MEC holds a Royal Society Research Professorship.

# Appendix 1:  Convergence and Small-mode approximations

Here we provide a sense for the computational cost of evaluating the STARM model. Our numerical method of choice transforms the unbounded domain $p > 0$ to a bounded domain $x = -1 + 2p/(1 + p)$ and then applies Chebyshev collocation. In Figure 30 and Figure 31, we show that for $\theta = 100$ and $c_v = 0.1$, calculations for the linear and nonlinear rheology appear to converge with as few as $N \sim 20$ and $N \sim 6$ collocation points, respectively.

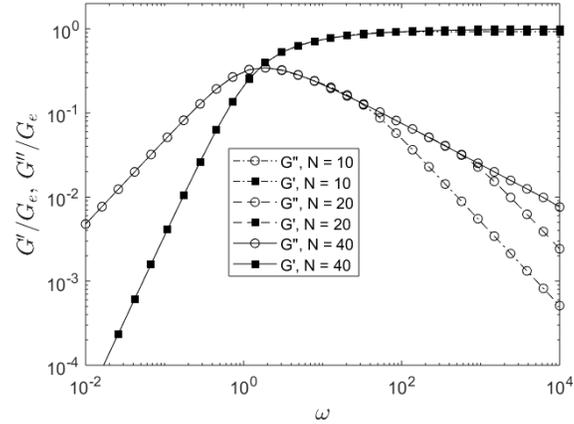

*Figure 30: Predictions for the linear rheology, $G'$, $G''$, as a function of the applied oscillation frequency $\omega$ for numerical evaluations of the STARM model with $N = 10, 20, 40$ evaluations in $p$.*

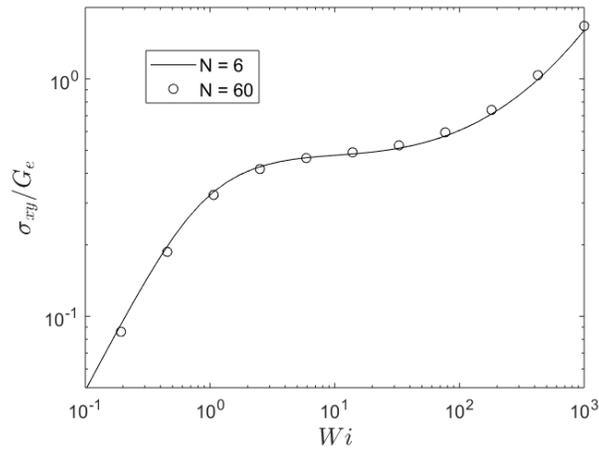

*Figure 31: Predictions for the steady state shear stress, $\sigma_{xy}$, as a function of the applied $Wi$ for numerical evaluations of the STARM model with $N = 6, 60$ evaluations in $p$. The calculations with $N = 6$ are very similar to the fully converged calculations of $N = 60$. Indeed, calculations are converged to within the line thickness by $N = 10$.*

With our numerical scheme, calculations are somewhat slower to converge for smaller $c_\nu$. However, for $c_\nu = 0$ the STARM model is equivalent to the Rolie Poly model with its constraint release parameter, $\beta$, also set zero [19]. It is also worth noting that even for $c_\nu \sim 0.1$, the nonlinear rheology of the Rolie Poly model (with an appropriately chosen value of $\beta$) is very similar to that of the STARM model. In Figure 32, we compare converged calculations of the STARM model ($\theta = 100, c_\nu = 0.1$) with calculations of the Rolie Poly model ($\theta = 100, \beta = 1$) over a wide range of $Wi = \dot{\gamma}\tau_0$.

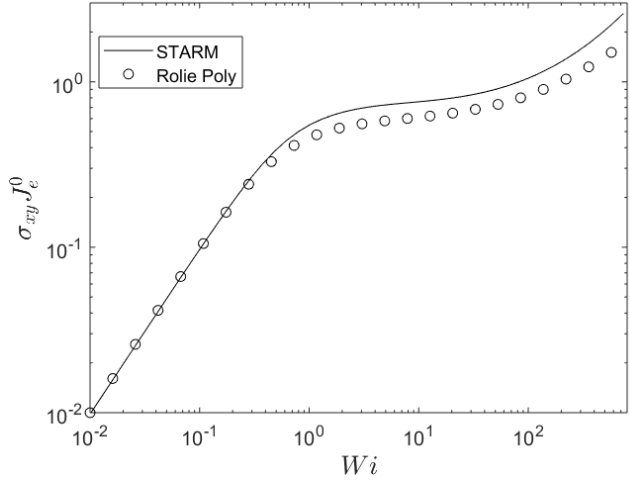

*Figure 32: Predictions for the steady state shear stress, $\sigma_{xy}$, as a function of the applied $Wi$ for numerical evaluations of the STARM model (solid line) and the Rolie Poly model (open circles). The steady shear rheology of STARM and Rolie Poly are similar over a very wide range of $Wi$ – the main difference is that the STARM model predicts slightly larger stresses for $Wi > 1$.*

## Appendix 2: Linear rheology of the 'toy' shuffling model for any $\zeta_{sh}$

The 'toy' shuffling model is introduced in section 2.1 as a simplified version of the full population balance description for reversible scission and end attack.

In its dimensionless units, the 'toy' shuffling model is an equation for the probability $P(t, s, z)$ that at time $t$ a tube segment in a chain of length $z$ and contour position $s$ was present at $t = 0$:

$$\frac{\partial}{\partial t} P = \frac{1}{z} \frac{\partial^2 P}{\partial s^2} - \zeta_{sh}^{-1}(P - \bar{P}) \qquad (100)$$

$$\bar{P}(t) = \int_0^\infty dz \, z e^{-z} \int_0^z ds \, P(t, s, z) \qquad (101)$$

$$P(t, s = \{0, z\}, z) = 0 \qquad (102)$$

$$P(t = 0, s, z) = 1 \qquad (103)$$

For the analysis that follows, it is important to recall that the complex modulus $G^*(\omega)$ is related to the Laplace transform of $\bar{P}$:

$$\tilde{\bar{P}}(q) = \int_0^\infty e^{-tq} \bar{P}(t) dt \qquad (104)$$

$$\frac{G^*(\omega)}{G_e} = i\omega \tilde{\bar{P}}(i\omega) \qquad (105)$$

Therefore we will not pursue a full solution to the 'toy' shuffling model – we are only interested in a simplified expression for the complex modulus. We begin by Laplace transforming:

$$\tilde{P}(q) = \int_0^\infty e^{-tq} P(t) dt \qquad (106)$$

$$q\tilde{P} - 1 = \frac{1}{z}\frac{\partial^2 \tilde{P}}{\partial s^2} - \zeta_{sh}^{-1}\left(\tilde{P} - \bar{\tilde{P}}\right) \qquad (107)$$

This can be re-arranged:

$$\frac{\partial^2 \tilde{P}}{\partial s^2} - A\tilde{P} = B \qquad (108)$$

$$A = z\left(\zeta_{sh}^{-1} + q\right) \qquad B = -z\left(q + \zeta_{sh}^{-1}\bar{\tilde{P}}\right) \qquad (109)$$

For simplicity, we shift the boundary conditions to:

$$\tilde{P}\left(s = \pm\frac{z}{2}\right) = 0 \qquad (110)$$

Note that because $A$ and $B$ have no $s$-dependence, we can solve by means of an integration factor:

$$\tilde{P} = -\frac{B}{A}\left[1 - \frac{\cosh\left(s\sqrt{A}\right)}{\cosh\left(\frac{z}{2}\sqrt{A}\right)}\right] \qquad (111)$$

We can now solve for $\bar{\tilde{P}}$ by integrating the expression for $\tilde{P}$ in equation (111) over the all contour positions and all chain lengths:

$$\bar{\tilde{P}} = C\left[1 + \zeta_{sh}^{-1}\bar{\tilde{P}}\right] \qquad (112)$$

$$\bar{\tilde{P}} = \left[\frac{1}{C} - \frac{1}{\zeta_{sh}}\right]^{-1} \qquad (113)$$

$$C \equiv \int_0^\infty dz \frac{ze^{-z}}{A/z}\left[1 - \frac{2}{z\sqrt{A}}\tanh\left(\frac{z}{2}\sqrt{A}\right)\right] \qquad (114)$$

Thus, we obtain our final result for the complex modulus:

$$\frac{G^*(\omega)}{G_e} = i\omega\hat{P}(i\omega) = i\omega\left[\frac{1}{C} - \frac{1}{\zeta_{sh}}\right]^{-1} \qquad (115)$$

$$C\int_0^\infty dz \frac{ze^{-z}}{A/z}\left[1 - \frac{2}{z\sqrt{A}}\tanh\left(\frac{z}{2}\sqrt{A}\right)\right] \qquad (116)$$

$$A = z\left(\zeta_{sh}^{-1} + i\omega\right) \qquad (117)$$

This is the result given in the main text, equations (96)-(98). Note also that this final result is equivalent to the final result for reptation and scission found in the 'Poisson Renewal' model of Granek and Cates [10], except that here we have excluded the possibility of a length-dependent shuffling time.

# Appendix 3:  Additional Discussion on Alignment and Attachment

In section 2.4, we presented rules for constructing tangent correlation tensors of product chains from the tangent correlation tensors of reactant chains.  A key component of those rules was that for any system in which chain ends are isotropically oriented (e.g. flexible polymers), any time two chains combine, the interior segments of the two chains will be totally uncorrelated with one another (except to the extent required by a finite segment length or persistence length).  We also assume this rule for systems in which chain ends are not isotropically oriented, provided the kinetics of end-attachment do not depend on the degree to which the reactant chain ends are aligned.  Whenever a system fails to meet one of these criteria, predicting the alignment of previously separate chains after they join is a more complicated task.

A natural point of objection to the proposed rules emerges when one considers a system of oriented rods – if all of the rods are oriented in the same direction, how could it be that two rods would join together to form an object for which the orientation of its 'top' and 'bottom' halves are uncorrelated? Consider an ensemble of rods with a uniform orientation.  When we think about a collection of rods oriented in this way, intuitively we assume that the system must have a preference for rods to attach in parallel (as opposed to anti-parallel) alignment.  Implicitly, we assume that this system arrived at its presently oriented and aligned state via the same kinds of reversible scission processes that will determine its future.  Here, however, we consider the oriented and aligned state to be a pre-set initial condition that may or may not reflect the future state of the system.

If the rod-like structures shown in Figure 33 have no preference to attach in parallel or anti-parallel, then we have assumed that, on average, there is no correlation in the alignment of the two rod-like sections subsequent to their attachment.  In the parallel attachment, the tangent vectors to the two rod segments are identical, and in the anti-parallel attachment they are opposite.  The *mean* tangent correlation between these two attached segments is therefore zero.

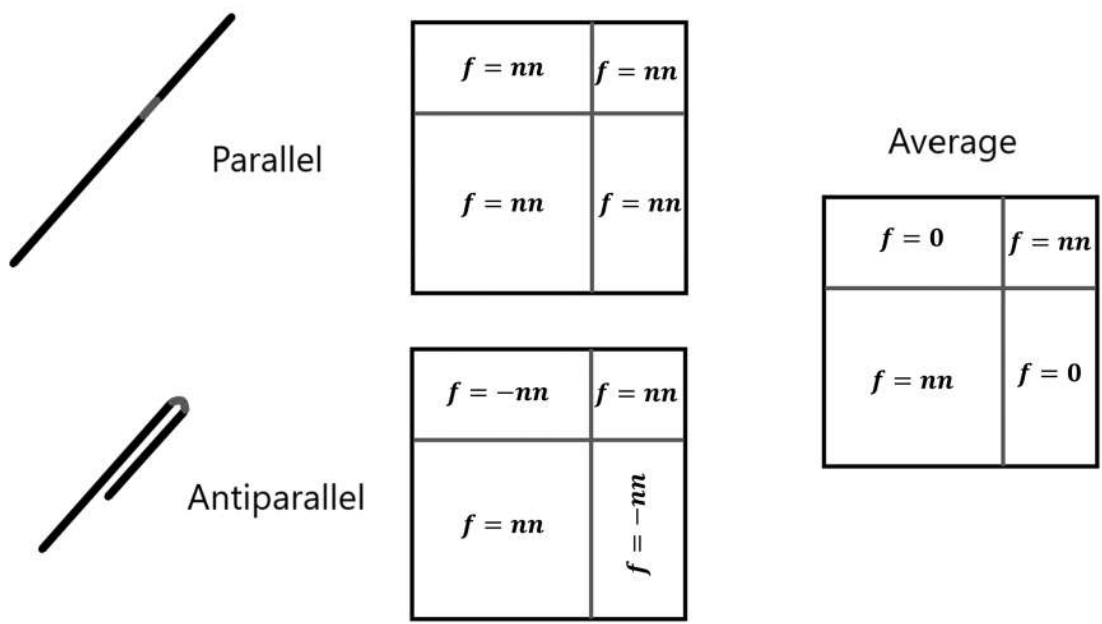

These arguments help explain why alignment and orientation of interior chain segments is not incompatible with our claim that the tangent vectors of two flexible chains are, on average, uncorrelated subsequent to their attachment. The generalization to semi-flexible chains (wherein we allow correlations at the equilibrium level) is natural but nonetheless approximate.

The notion of uncorrelated attachment is correct for infinitely flexible chains, but the generalization to semi-flexible chains in equation (79) is approximate. Therefore, we should also establish that the approximation does not have a large effect on the final rheological predictions, at least for well-entangled semi-flexible chains for which the persistence length is much smaller than the typical chain length. For semi-flexible chains, the precise rule for alignment subsequent to attachment is not known but upper and lower bound estimates are. First, as a lower bound, the correlations that exist between two recently attached chains can be no less than the correlations that would exist between those chains were they to attach in the equilibrium state (this is the rule that we have used for the STARM model, c.f. equation (76)). Second, shuffling tube segments randomly through the molecular weight distribution (by whatever reaction pathway one chooses) cannot increase the average alignment between tube segments of any separation – it must reduce alignment or (as an absolute upper bound) leave it unchanged. This upper bound is the rule that's enforced if one removes (79) from the STARM model altogether. If the upper bound and lower bound estimates led to wildly varying rheological predictions, one would need to cautiously assess which (if either) of the two was likely to be more accurate. Fortunately, however, choosing the upper bound or the lower bound (or anything in between) makes virtually no difference to the linear and nonlinear rheological predictions of STARM in the parameter regimes explored in the current paper.